\documentclass[11pt]{article}

\usepackage{amsmath}
\usepackage{amssymb}
\usepackage{epsfig}
\usepackage{psfrag}
\usepackage{graphicx}
\usepackage{here}

\normalsize

\begin{document}

\begin{titlepage}
\title{Emergence of intense jets and Jupiter Great Red Spot as maximum entropy structures.}

\author{F. BOUCHET{$^1$} and J. SOMMERIA{$^2$}  \\
\\
{\small
\parbox{11.5cm}{$^1$\parbox[t]{11cm}{UMR 5582, Institut Fourier, BP 74, 38402
    Saint Martin d'H\`eres Cedex, France}\\
\vskip 0.3cm
$^2$\parbox[t]{11cm}{CNRS, LEGI/Coriolis, 21 av. des Martyrs, 38 000 Grenoble, France}
}}
}
\date{\today}

\end{titlepage}

\maketitle

\begin{abstract}
We explain the emergence and robustness of intense jets in highly turbulent planetary atmospheres, like on Jupiter, by a general approach of statistical mechanics of potential
 vorticity patches. The idea is that potential vorticity mixing leads to the formation of a steady organized coarse grained flow, corresponding to the statistical equilibrium
state. Our starting point is the quasi-geostrophic 1-1/2 layer model, and we  consider the relevant limit of a small Rossby radius of deformation. Then narrow jets
are obtained, scaling like the Rossby radius of deformation. These jets can be
 either zonal, or closed into a ring  bounding a vortex. Taking
into account the effect of the beta effect and a sublayer deep shear flow, we
 predict an organization of the turbulent atmospheric layer into an oval-shaped 
vortex amidst a background shear. Such an isolated vortex is centered over an
extremum  of the equivalent topography (determined by the deep shear flow and
 beta-effect). This prediction is in agreement with analysis of wind data in major
Jovian vortices (Great Red Spot and Oval BC).


\end{abstract}

\section{Introduction}
Atmospheric and oceanic flows are often organized into narrow jets. They can 
zonally flow around 
the planet like the jet streams in the 
earth stratosphere, or the eastward jet at 24° in the northern hemisphere of 
Jupiter ( Maxworthy 1984). Jets can alternatively organize into rings, forming vortices, like the 
rings shed 
by the meandering of the Gulf-Stream in the western Atlantic Ocean. The flow field in Jupiter most 
famous feature, the Great Red Spot,
 is an oval-shaped jet, rotating in
 the anticyclonic direction and surrounding an interior area with a weak
 mean flow ( Dowling and Ingersoll 1989), see figure 1(a). Robust cyclonic vortices have a similar jet structure ( Hatzes et al 1981 ), see figure 1(b).

Such jets and vortices are in a turbulent surrounding, and the persistence of their strength
 and concentration in the presence of eddy 
mixing is intriguing. The explanation proposed in this paper is based on a statistical
 mechanical approach: the narrow jet or vortex appears as the most probable state of the flow
 after a turbulent mixing of potential vorticity, 
taking into account constraints due to the quantities conserved by the dynamics, especially 
energy. Such a
 statistical theory has been first proposed for the two-dimensional Euler equations by
 Kuz'min (1982), Robert (1990)
, Robert and Sommeria (1991), Miller (1990). See Brandt et al (1999) for a recent 
review and discussion. This theory predicts an organization of two-dimensional turbulence into a
 steady flow (with fine scale, 'microscopic' vorticity fluctuations). Complete
vorticity mixing is prevented
by the conservation of the energy, which can be expressed as a constraint in the accessible 
vorticity fields. A similar, but quantitatively different, organization had been previously obtained
 with statistical mechanics
of singular point vortices with the mean field 
approximation, instead of continuous vorticity fields (Onsager 1949, Joyce and Montgomery 1973). 

Extension to the quasi-geostrophic (QG) model has been discussed by 
Sommeria et al (1991), Michel \& Robert (1994), Kazantsev Sommeria and Verron (1998). This model describes a shallow water system with a
weak vorticity in comparison with the planetary vorticity (small Rossby number), such that the
pressure is in geostrophic balance, and the corresponding free surface deformation is supposed small
in comparison with the layer thickness. For Jupiter
the free surface would be rather at the bottom of the active atmospheric
 layer, floating on a denser fluid, as discussed by Dowling and Ingersoll (1989), see
 Dowling (1995) for a review. The gradient of planetary vorticity is accounted by a beta-effect. An additional
beta-effect, depending on the latitude coordinate $y$, is introduced to represent the influence on the 
active atmospheric layer of a steady zonal flow in the deep interior, as discussed by Dowling and Ingersoll (1989). 

The free surface deformability, representing the strength of the density stratification, is
 controlled by
the Rossby radius of deformation $R^*$. The two-dimensional Euler equation is recovered in the
 limit of very strong stratification for which $R^* \to \infty$. We shall consider in this paper the 
opposite limit of weak stratification for which $R^*$ is much smaller than the scale of the
 system $L$. This limit is appropriate for large scale oceanic  currents, as the radius of
 deformation is typically 10-100 km. For Jupiter, $R^*$ is estimated to be in the range
 500-2500 km, while the Great Red Spot extends over 20,000 km in longitude,
 and 10,000 km in latitude, so 
the limit $R^*/L\to 0$ seems
relevant. We show that in this limit the statistical equilibrium is made of quiescent zones with
 well mixed uniform potential vorticity, bounded by jets with thickness of order $R^*$. This provides
therefore a general justification of jet persistence. Some of the ideas used have been already sketched in 
Sommeria et al (1991), but we here provide a systematic derivation and thorough analysis.

In
 principle, the Quasi Geostrophic approximation breaks 
down for scales much larger than the radius of deformation, so that 
the limit $R^*/L\to 0$ seems inconsistent with the QG approximation. However the relevant scale is
the jet 
width, which remains
of order $R^*$, so that the Quasi Geostrophic approximation remains valid in this limit. This point has been discussed
by Marcus (1993) for the Great Red Spot, which he supposes to be a uniform Potential Vorticity ( PV ) spot
 surrounded by a uniform Potential Vorticity
background ( we here justify this structure as the result of Potential Vorticity mixing with 
constraints on the conserved quantities ). Analyzing wind data in the Great Red Spot, Dowling
 \& Ingersoll (1989) concluded that the QG approximation is good up within typically
  30\% error, which is reasonable to a first approximation. Statistical mechanics of the 
more general shallow water
system (to be published), predicts a similar jet structure. The present Quasi Geostrophic
 results therefore provide a good description as a first approximation. 

We first consider the case without beta-effect in section 2. We furthermore assume
periodic boundary conditions (along both coordinates) in this section to avoid 
consideration of boundary effects. Starting from some initial condition with patches of uniform
PV, we find that these patches mix with uniform density (probability) in two sub-domains, with
 strong 
density gradient at the interface, corresponding to a free jet. The coexistence of 
the two sub-domains can be interpreted as an equilibrium between 
two thermodynamic phases. We find that the interface has a free energy per unit of length, and its 
minimization leads to a minimum length at equilibrium. This results
in a constant radius of curvature, in analogy with surface tension effects in thermodynamics,
 leading to spherical bubbles
or droplets. The range of the vortex interaction is of the 
order $R^*$, therefore becoming very small in the limit of small
 radius of deformation, so the statistical equilibrium indeed behaves like 
in usual thermodynamics with short range molecular interactions. This contrasts
 with the case of Euler equation, with 
long range vortex interactions, analogous to gravitational effects (Chavanis Sommeria and Robert 1996, Chavanis 1998).

 Figure \ref{phase} summarizes the calculated equilibrium states, depending on the total
energy and a parameter $B$ representing an asymmetry between the initial Potential Vorticity
 patch areas, before the mixing process. We obtain straight jets for a weak asymmetry and circular jets for
 higher asymmetry. Such circular a jet reduces to an axisymmetric vortex, with radius of order $R^*$, in the 
limit of low energy.

We discuss the influence of the beta-effect or the deep zonal flow in section 3. The 
channel geometry,
representing a zonal band periodic in the longitude $x$ is appropriate for that
study. With the usual
beta-effect $\beta y$, linear in the transverse coordinate $y$, statistical
 equilibrium is, depending on the initial parameters, a zonal flow, or a meandering eastward jet, or a uniform velocity $v_m=R^2\beta$ whose induced free surface slope cancels the beta-effect ( uniformization of Potential Vorticity ) on which circular vortices can coexist.

For more general beta-effects, due to the deep zonal flow, we find that the jet curvature depends on latitude
$y$. In particular a quadratic beta effect $ay^2$ leads to oval-shape jets, similar to the Great 
Red Spot. Using the determination of the sublayer flow
  from Voyager data by Dowling and Ingersoll (1989), we show in section 4, that such a
 quadratic effective beta-effect is indeed a realistic model for Jupiter atmosphere in the latitude 
range of the Great Red Spot and the White Ovals, the other major coherent vortices on 
Jupiter. Using these data on beta-effect, as well as the shear in the zonal flow at the latitude 
of the Great Red Spot, the jet width and its maximum velocity, we deduce all the parameters of our model.

 \section{The case with periodic boundary conditions}\label{per}

\subsection{The dynamical system}

We start from the barotropic Quasi Geostrophic (QG) equation : 

\begin{equation}
{\frac{\partial q}{\partial t}}+{\bf v\cdot \nabla }q=0
\label{QG}
\end{equation}
\begin{equation}
 \quad q=-\Delta \psi +
\frac{\psi}{R^2} - h(y)
\label{dir}
\end{equation}
\begin{equation}
\label{u}
{\bf v} =-\hat {{\bf z}}\wedge \nabla \psi 
\end{equation}
where $q$ is the potential vorticity ( PV ), advected by the non-divergent velocity ${\bf v}$, $\psi$ is the stream function, $R$ is the internal Rossby deformation radius between the
layer of fluid under consideration and a deep layer unaffected by the
dynamics. $x$ and  $y$ are respectively the zonal and meridional
coordinates ( $x$ is directed eastward and $y$ poleward ). The term $h(y)$ represents the combined effect of the
planetary vorticity gradient and of a given stationary
zonal flow in the deep layer, with stream
 function $\psi_d(y)$: $h(y)=-\beta y+\psi_d/R^2$. This deep flow induces a constant deformation of the free
surface, acting like a topography on the active layer. We shall therefore call $h(y)$
 the 'topography', and study its influence 
in section 3. Let us assume $h(y)=0$ in this section.
We define the QG equations (\ref{QG},\ref{dir}) in the non-dimensional square
$D = [-\frac{1}{2},\frac{1}{2}]^{2}$. $R$ is then the ratio of the internal Rossby deformation
radius $R^*$ to the physical scale of the domain $L$. 

Let $\langle f \rangle \equiv \int_D fd^{2}{\bf r}$ be the average of $f$ on $D$ for any 
function $f$. Physically,
as the stream function $\psi$ is related to the geostrophic pressure,
$\langle \psi \rangle$ is proportional to the
mean height at the interface between the fluid layer and
the bottom layer, and due to the mass conservation it must be constant (Pedlosky 1987). We make the choice
\begin{equation}\label{psi0}\langle \psi \rangle = 0
\end{equation}
without loss of generality. 

 The total circulation is
 $\langle q \rangle = \langle - \Delta \psi + \psi / R^2 \rangle = \langle \psi /R^2 \rangle $ 
due to the periodic boundary conditions. Therefore 
\begin{equation}
\label{qzero}
\langle q \rangle = 0
\end{equation} 
We note that the Dirichlet problem (\ref{dir}) on $D$ with
periodic boundary conditions has a unique solution $\psi$ for a given
PV field. 

Due to the periodic conditions for $\psi$, the linear momentum is also equal to 0,
\begin{equation}
\label{vitessemoy}
\langle  {\bf v} \rangle=0
\end{equation} 
 
The energy 
\begin{equation}
\label{ene}
E={\frac{1}{2}}\int_D q\psi d^{2}{\bf r}={\frac{1}{2}}\int_D
[\ ({\bf \nabla}\psi)^2 + \frac{\psi ^2}{R^2} \ ] d^{2}{\bf r}
\end{equation}  
is conserved ( we note that the first term in the right hand side of (\ref{ene}) is
the kinetic energy whereas the second one is the gravitational
available  potential energy ). 
 
The integrals
\begin{equation}
\label{cas}
C_f(q) = \int_D f(q)d^{2}{\bf r}
\end{equation}
for any continuous function $f$ are also conserved, in particular the
different moments of the PV. In the case of an
initial condition made of a finite number of PV levels,
 the areas initially occupied
by each of these levels is conserved, and this is equivalent 
to the conservation of all the constants of motion (\ref{cas}).

\subsection{The statistical mechanics on a two PV levels configuration.}

\subsubsection{The macroscopic description.}

The QG equations (\ref{QG}) (\ref{dir}) are known to develop
very complex vorticity filaments.  Because of the rapidly increasing
amount of information it would require, as time goes on, a
deterministic description of the flow for long time is both
unrealistic and meaningless.
The statistical theory adopts a probabilistic description for the
vorticity field. 
The statistical equilibrium depends on the energy and of the global probability
 distribution of PV levels. Various previous studies
 (Sommeria \& al 1991), (Kazantsev \& al 1998) indicate
 that a model with only two PV levels provides a good approximation in many 
cases. The determination of the statistical equilibrium is then simplified as it depends only on the energy, on the two PV levels, denoted  $q = a_1$ and $q =
a_{-1}$ and on their respective areas  $A$ and $(1 - A)$ in $D$. The number of free parameters can be further 
reduced by appropriate scaling. Indeed a change in the time unit permits to define the PV levels
 up to a multiplicative constant, and we choose for the sake of simplicity : 
\begin{equation}
\label{csttemps}
\frac{a_1 - a_{-1}}{2} = 1
\end{equation}
and define the non-dimensional parameter $B$ as :
\begin{equation}
\label{B}
B \equiv \frac{a_1+a_{-1}}{2}
\end{equation}
The condition (\ref{qzero}) of zero mean PV imposes that $a_1 A+a_{-1}(1-A)=0$. This means that $a_1$ and 
$a_{-1}$ must be of opposite sign and, using (\ref{csttemps})
and (\ref{B}), $A=(1-B)/2$.  The distribution of PV levels is therefore fully 
characterized by the single asymmetry parameter $B$, which takes values between -1 and +1. The
 symmetric case of two PV patches
 with equal area $A=1/2$ corresponds to $B=0$, while 
the case of a patch with small area (but high PV, 
such that $\langle \ {q} \  \rangle = 0 $) corresponds to $B\to 1$. Note 
that we can restrict the discussion to $B \ge 1$ as the QG system is symmetric by a
 change of sign of the PV.  

The two PV levels mix due to turbulent effects, and the resulting state is 
locally described by the local probability (local area
proportion)
$p({\bf r})$ to find the first level at the location ${\bf r}$. The probability to find the complementary
PV level $a_{-1}$ is $1-p$, and the locally averaged PV at each
point is then 
\begin{equation}
\label{PVmean}
\overline{q}({\bf r}) = a_1p({\bf r}) + a_{-1}(1-p({\bf
  r})) = 2\left(p-\frac{1}{2}\right) + B 
\end{equation}
where the second relation is obtained by using (\ref{csttemps}) and (\ref{B}).

As we consider the evolution of two PV patches, the conservation of all
 invariants (\ref{cas}) is equivalent to the conservation of the area $A$ of the patch with PV 
value $a_1$ ( the area of the other PV level $a_{-1}$ being $1-A$ ).
The integral of $p$ over the 
domain must be therefore equal to the initial area $A$ ( the patch with PV level 
$a_1$ is mixed but globally conserved ),
\begin{equation}
\label{PVcons}
A\equiv \frac{1-B}{2}=\int_D p({\bf r})d^{2}{\bf r} 
\end{equation}
 
As the effect of local PV fluctuations is filtered out by
integration, the stream function and the velocity field are fully
determined by the locally averaged PV $\ \overline{q} \ $ as the
solution of 
\begin{equation}
\label{dirmean}
\overline{q} =-\Delta \psi +
\frac{\psi}{R^2}\ ; \  \psi \   \ periodic \  
\end{equation}
$$ and \ \ \ {\bf v} =-\hat {{\bf z}}\wedge \nabla \psi $$
Therefore the energy is also expressed in
terms of the field $\overline{q}$ : 
\begin{equation}
\label{Econs} 
E = {\frac{1}{2}}\int_D
\left[\ ({\bf \nabla}\psi)^2 + \frac{\psi ^2}{R^2} \ \right] d^{2}{\bf r} = \frac{1}{2}\int_D \overline{q}\psi d^{2}{\bf r}  
\end{equation}
Here the energy of the 'microscopic' PV
 fluctuations has been neglected (replacing $q$ by $\overline{q}$), as justified in
 the case of Euler equation
by Robert and Sommeria (1991). Indeed, considering a 'cutoff' for the microscopic
 fluctuations much smaller
than $R$, the small scale dynamics coincides with the Euler case.

The central result of the statistical mechanics of the QG equations
(\ref{QG},\ref{dir}) is that, under an ergodic hypothesis, we expect the long time
dynamics to converge towards the Gibbs states defined by maximizing the mixing entropy 
\begin{equation}
\label{ent}
S = - \int_D  [ \  p({\bf r})\ln p({\bf r}) + (1-p({\bf r}))\ln (1-p({\bf
  r})) \ ]d^{2}{\bf r} 
\end{equation}
under the constraints of the global PV distribution (\ref{PVcons}) and energy
(\ref{Econs}). It can be shown that the microscopic states satisfying the constraints 
given by the conservation laws are overwhelmly concentrated
near the Gibbs state, which is therefore likely to be reached after a
complex flow evolution. A good justification of this statement is obtained by the 
construction of
converging sequences of approximations of the QG equation (\ref{QG},\ref{dir}), in finite
dimensional vector spaces, for which a Liouville theorem holds. This
is a straightforward translation of the work of Robert
(1999) for 2D Euler equations. The sequence of such Liouville measures
has then the desired concentration properties as
(\ref{QG},\ref{dir}) enters in the context considered in 
Michel \& Robert (1994).

\subsubsection{The Gibbs states}

Following Robert \& Sommeria (1991), we seek maxima of the entropy (\ref{ent}) under
 the constraints (\ref{PVcons}) and 
(\ref{Econs}). To account for these 
 constraints, we introduce 
two corresponding Lagrange multipliers, which we denote $2\alpha$ and $-C/R^2$
 for convenience in future calculations.
Then the first variation of
the functionals satisfies : 
$$
\label{delF}
\delta S - 2\alpha \delta A + \frac{C}{R^2}\delta E = 0
$$
for all variations $\delta p$ of the probability field $p$. After straightforward
 differentiation  we obtain:
$$
\delta S = - \int_D  [ \  \ln p - \ln (1-p) \ ]\delta pd^{2}{\bf r}\;\;,
\;\;\delta A = \int_D \delta pd^{2}{\bf r} 
$$
\begin{equation}
\label{deltaE}
\delta E = \int_D \psi \delta \overline{q} d^{2}{\bf r} = \int_D 2\psi \delta pd^{2}{\bf r} 
\end{equation}
where the expression of $\delta E$ has been obtained by integrating by part 
and expressing $\overline{q}$ by
 (\ref{PVmean}). Then we can write
 the first variation 
 under the form $\int_D [ \ -\ln p + \ln (1-p) - 2\alpha +2C\psi/R^2\ ]\; \delta p\;d^{2}{\bf r}$
 which must vanish for any small variation $\delta p$. This implies that the integrand
 must vanish, and yields the 
 equation for the optimum state:
\begin{equation}
 p = \frac{1-\tanh (\alpha - \frac{C\psi}{R^2})}{2},
\label{probagibbs}
\end{equation}
and using (\ref{PVmean}) and (\ref{dirmean}), the partial differential equation   
\begin{equation}
q=-\Delta \psi +\frac{\psi}{R^2} = B \ - \ \tanh \left(\alpha - \frac{C\psi}{R^2}\right)
\label{gibbs}
\end{equation} 
determining the Gibbs states (statistical equilibrium). From now on we forget the $q$
over-line for the locally averaged PV and refer to it as the PV.

Therefore, we have shown that for any solution of the variational problem, two constants $\alpha $ and
$C$ exist such that $\psi$ satisfies (\ref{gibbs}). Conversely
it can be proved that for any such two constants, a solution to equation
(\ref{gibbs}), in general not unique, always exists. Then $p$
associated with one of these solutions by (\ref{probagibbs}) is a critical
point of the 'free energy'
$-S(p) + 2\alpha A(p) - \frac{C}{R^2}E(p)$ (i.e. its first variation vanishes).
 Then the Lagrange multipliers are not given but have to be calculated by prescribing
 the constraints (\ref{PVcons}) and 
(\ref{Econs}) corresponding to the two parameters $B$ and $E$ respectively, given by
the initial condition. Furthermore, among the states
 of given energy $E$ and asymmetry parameter $B$, we shall select the actual maxima.

Finally, let us find a lower bound for the parameter $C$ of the Gibbs
states with non-zero energy (i.e. $\psi$ is not constant on $D$). 
 Multiplying (\ref{gibbs}) by $- \Delta \psi$, integrating by part
 and defining  $f(C\psi ) \equiv B - \tanh(\alpha - \frac{C\psi}{R^2})$,
 we obtain :
$$
 C = \frac{\int_D \left( (\Delta \psi)^2 + \frac{1}{R^2}({\bf
  \nabla}\psi)^2 \right)d^{2}{\bf r}}{\int_D - f^{\prime}(C \psi)({\bf
  \nabla}\psi)^2 d^{2}{\bf r}}
$$
From which, using $0 < -  f^{\prime
  }(C\psi) \leq \frac{1}{R^2}$ it follows that when $\psi$ is not constant : 
\begin{equation}
\label{estimation}
C > 1
\end{equation}

\subsection{The limit of small Rossby deformation radius}

As suggested by oceanographic or Jovian parameters, we seek solutions for
the Gibbs states equation in the limit of a small ratio between the
Rossby deformation radius and the length scale of the domain : $R <<
1$ with our non-dimensional coordinates \footnote{ Modica (1987) considered the minimization of the functional $E_\epsilon (u) = \int_{\Omega} [ \epsilon \left({\bf \nabla}u\right)^2 + W_0(u)]d{\bf x}$ with the constraint $\int_{\Omega} u({\bf x})d{\bf x} = m$ in the limit $\epsilon \to 0^+$ where $W_0$ is a real function with two relative minima. He proved, in a mathematical framework, working with bounded variations functions, that if $(u_{\epsilon})$ are solutions
of this variational problem, for any subsequence of $(u_{\epsilon})$ converging in $L^1(\Omega)$ as $\epsilon \to 0$, this subsequence converge to a function $u_0$ which takes only the values where $W_0$ reaches its minima ; with the interface between the corresponding subdomains having minimal area ( See Modica (1987) for a precise statement ). 

 We note that the Euler equation of this variational problem may be the same as the  Gibbs States equation (\ref{gibbs}) for a convenient choice of $W_0$. However as the variational problem  itself is different this beautiful result cannot be used in our context.}

\subsubsection{The uniform subdomains}

\label{secsub}

Then we expect that the Laplacian in the Gibbs states equation (\ref{gibbs}) can 
be neglected with respect to $\psi /R^2$, except possibly in transition regions
of small area. This transforms (\ref{gibbs}) into the algebraic equation : 

\begin{equation}
q = \frac{\psi}{R^2} = B \ - \ \tanh \left(\alpha - \frac{C\psi}{R^2}\right)
\label{alg}
\end{equation}

Depending on the parameters, this equation has either one, two or
three solutions, denoted $\psi_{-1}, \psi_{0}$ and $\psi_{1}$ in
increasing order (see figure \ref{conint} ). The case with a single
solution would correspond to a uniform $\psi$, which should be equal
to 0 due to the condition $\langle \psi \rangle = 0$. This is only
possible for $E=0$. Otherwise, we have therefore two or three solutions, with
 different solutions occurring in subdomains. This condition of multiple solutions
 requires that the maximum slope for the right hand side of (\ref{alg}) must be greater
 than $1/R^2$ ; this is always realized due to the inequality (\ref{estimation}). Furthermore
 $\alpha $ must be 
in an interval centered in $CB$ ( $\alpha=CB$ in the symmetric case of figure \ref{conint} ).

At the interface between two constant stream function subdomains, a
strong gradient of $\psi$ necessarily occurs, corresponding to a jet along the
interface. These jets give first order contributions to the entropy and energy, but let us 
first describe the zero order problem.  
Suppose that $\psi$ takes the value $\psi_1$ ( resp $\psi_{-1}$ ) in subdomains 
of total area $A_1$ ( resp $A_{-1}$ ). The reason why we do not select the value $\psi_0$ 
will soon become clear. Using (\ref{PVmean}) we conclude that the probability $p$ takes 
two constants values $p_{\pm 1}$ in their respective subdomains. The two areas $A_{\pm1}$ 
( measured from the middle of the jet ) are complementary such that : 
\begin{equation}
\label{AA}
A_{1} + A_{-1} = 1
\end{equation}
Furthermore the constraint (\ref{psi0}) of zero domain average for $\psi$ implies at zero order,
\begin{equation}
\label{PA}
\psi_{1}A_{1} + \psi_{-1}A_{-1} = 0,  
\end{equation}
or equivalently, using $q_{\pm 1} = \psi_{\pm 1}/R^2$, (\ref{PVmean}) and (\ref{AA}) :
\begin{equation}
\label{pA}
2A_{1}\left( p_{1}-\frac{1}{2}\right) + 2(1-A_{1})\left(p_{-1}-\frac{1}{2}\right) = - B
\end{equation} 
This can be obtained as well from the constraint on the (microscopic) PV patch 
area (\ref{PVcons}). The energy inside the subdomains
 reduces to the potential term $\psi^2/2R^2$, since
velocity vanishes. This area energy $E_A$ can be computed in terms 
of $p_{\pm 1}$ using $q_{\pm 1} = \psi_{\pm 1}/R^2$ and (\ref{PVmean}) : 
\begin{equation}
\label{EA}
\begin{split}
E_A &\equiv \frac{\psi_{1}^2A_{1} +  \psi_{-1}^2A_{-1}}{2R^2} =  A_1e(p_1) + (1-A_{1})e(p_{-1}) \\
\text{with} \ \ e(p) &\equiv R^2\left(2\left(p-\frac{1}{2} \right)^2 +  2B\left(p-\frac{1}{2} \right) + \frac{B^2}{2}\right)
\end{split}
\end{equation}
There is also an energy in the jet at the interface of subdomains, but it
 is small with respect to $E_A$. Indeed the velocity in the jet, of width $R$, is 
of order $(\psi_1-\psi_{-1})/R\sim \psi/R$, and the corresponding integrated kinetic 
energy is of order $\psi^2/R$. This
is small in comparison with the area energy $E_A$ (mostly potential) of order $\psi^2/R^2$. A precise calculation
will confirm this estimate in next sub-section.

We need to determine three unknown, the area $A_1$ and the probabilities $p_{\pm 1}$ of the 
PV level $a_1$ in each subdomain, while the constraints (\ref{pA}) and (\ref{EA}) provide 
two relations. An additional relation will be given by entropy maximization. As we neglect the
jet area, the entropy reduces at order zero to the area entropy : 
\begin{equation}
\label{SA}
\begin{split}
S_A &\equiv A_1s(p_1) + (1-A_{1})s(p_{-1}) \\
\text{with} \ \ \ s(p) &\equiv -p\log p - (1-p)\log (1-p)
\end{split}
\end{equation}
Thus the zero order problem corresponds to the maximization of
 the area entropy (\ref{SA}) with respect to the 3 parameters $p_{\pm 1}$ and $A_{1}$, under
 the 2 constraints (\ref{pA}) and (\ref{EA}). A necessary condition for a solution of this
 variational problem is the existence of two Lagrange parameters $\alpha_0$ and $C_0$, associated
 respectively with the circulation constraint (\ref{pA}) and with the energy constraint (\ref{EA}),
such that the first variations of the total free energy,
\begin{equation}
\label{FA}
F_A \equiv -S_A - \frac{C_0}{R^2}E_A + \alpha_0 \frac{\langle \psi \rangle}{R^2},\end{equation}
vanish. Let us calculate $F_A$ using (\ref{pA}) and (\ref{EA}):
\begin{equation}
\begin{split}\label{fp}F_A &=  A_{1}f(p_{1}) + (1-A_{1})f(p_{-1}), \\
  \text{with} \ \  f(p) &\equiv   -s(p) - 2C_0\left(p - \frac{1}{2}\right)^2
- 2\left(C_0B - \alpha_0\right) \left(p - \frac{1}{2}\right) - \frac{C_0B^2}{2} - \alpha_0 B.
\end{split}
\end{equation} 
The vanishing of the variations with respect to $p_{1}$ and $p_{-1}$ gives that 
$f(p_{\pm 1})$ are local minima of the free energy $f(p)$. It is easily proven
that the function $f$ has two local minima and one local maximum ( for $C_0 > 1$ 
and $(C_0B - \alpha_0)$ small enough ) ( see figure \ref{figenlibre}). The local
 maximum is achieved for $p_0$ corresponding to the value $\psi_0$. It is the 
reason why it has not been taken into account in this analysis. In addition, the
 vanishing of the first variations with respect to the area $A_1$ imposes the free
 energies $f(p_{\pm 1})$ in the two subdomains to be equal. This is like the
 condition of thermodynamic equilibrium for a chemical species shared by two coexisting
phases.

In the expression (\ref{fp}) 
of $f(p)$, the entropy term $s(p)$ is symmetric with respect to $p = \frac{1}{2}$, as 
well as the quadratic term. Therefore if the linear term in $\left(p-\frac{1}{2}\right)$
 vanishes the two maxima are equal, with $p_{\pm 1}$ symmetric with respect to
 $\frac{1}{2}$. The addition of a linear term obviously breaks this condition of 
two equal maxima, so the coefficient of the linear term must vanish, thus : 
\begin{equation}
\label{alpha0}
\alpha_0 = C_0 B. 
\end{equation}
Since $p_{\pm 1}$ are symmetric with respect to $\frac{1}{2}$, we introduce the parameter $u$ by :    
\begin{equation}
\label{p}
p_{\pm 1} = \frac{1}{2}(1 \pm u).
\end{equation}
Using (\ref{PVmean}),(\ref{pA}) we deduce:
\begin{equation}
\label{psiu}
\psi_{\pm 1} = R^2(B\pm u)
\end{equation}
From (\ref{PA}) we state that the two constant stream
function (\ref{psiu}) have to be of opposite sign, so that $u > |B|$.
Introducing (\ref{p}) in the circulation constraint (\ref{pA}), and using (\ref{AA}), we get :
\begin{equation}
\label{aire}
A_{\pm 1} = \frac{1}{2}\left(1 \mp \frac{B}{u}\right). 
\end{equation}
Using these results, the energy (\ref{EA}) becomes
\begin{equation}
\label{eneu}
E \simeq E_A = \frac{R^2}{2}(u^2 - B^2)
\end{equation}
This relates the parameter $u$ to the given energy $E$ and asymmetry parameter $B$. 
Finally the condition that $f(p_{\pm 1})$ are maxima of $f$ leads to :
\begin{equation}
\label{uu}
u = \tanh (C_0u), 
\end{equation}
which determines the 'temperature' parameter $C_0$, as represented in figure \ref{figu}. Therefore all the quantities are 
determined from the asymmetry parameter $B$ and
from the parameter $u$, related to the energy by (\ref{eneu}).
\\

In the limit of low energy, $ u \rightarrow |B| $, when for
instance $B>0$, then $A_{1}$ goes to zero, so that $\psi_{-1}$ tends to occupy the
 whole domain. This state is the most mixed one compatible with the 
constraint of a given value of $B$ (or equivalently a given initial patch area $A=(1-B)/2$). In 
the opposite limit $ u \rightarrow 1 $, we see from (\ref{psiu})
that in the two subdomains $q
 = \psi /R^2$ tends to the two initial PV levels $a_{1} = 1 + B$
and $a_{-1} = -1 + B$. Thus, this state is an unmixed state. It
achieves the maximum possible energy $E = \frac{R^2}{2}(1 - B^2)$  under the
 constraint of a given patch area. We conclude that the parameter $u$, or the related 'temperature' $C_0$, linked with
 the difference between the energy and the maximum accessible energy for the two given
 initial PV levels, characterizes the mixing of these two PV levels. We shall call $u$ the segregation parameter, as it quantifies the segregation of the PV level $a_1$ ( or its complementary $a_{-1}$ ) between the two phases. \\

Let us now study the interface between the subdomains. 

\subsubsection{Interior Jets}

\label{jetequation}

At the interface between two constant stream function subdomains, a
strong gradient of $\psi$ necessarily occurs, corresponding to a jet along the
interface. To study these jets, we come back to the Gibbs state equation (\ref{gibbs}). We 
expect the Lagrange parameters $\alpha$ and $C$ to be close to the zero order 
parameters $\alpha_0$ and $C_0$, computed in the previous sub-section, so we use 
$\alpha = \alpha_0$ and $C =C_0$ to calculate the jet structure.
In such a jet, we cannot neglect the Laplacian term in
(\ref{gibbs}), but a 
boundary layer type approximation can be used: we neglect the
tangential derivative with respect to the derivative along the coordinate normal to the
interface, $\zeta $. Accordingly, we neglect the inverse of the curvature radius of the jet with 
respect to ${1}/{R}$.

Thus, from the Gibbs states equation (\ref{gibbs}), using (\ref{alpha0}), we 
deduce the jet equation : 
\begin{equation}
\label{jet1}
- \frac{d^2\psi}{d\zeta ^2} +\frac{\psi}{R^2} = B \ - \ \tanh \left(C_0\left(B - \frac{\psi}{R^2}\right)\right)
\end{equation}
 As the stream function depends only on the normal coordinate $\zeta$, the 
velocity is tangent to the interface, forming a jet with a typical width scaling like
$R$. We thus make the change of variables defined by :
\begin{equation}
\label{var}
\tau \equiv \frac{\zeta}{R} \ ; \ \phi \equiv -B + \frac{\psi}{R^2}, 
\end{equation}
leading to the rescaled jet equation:
\begin{equation}
\label{jet}
 \frac{d^2\phi}{d\tau ^2} = -\tanh (C_0\phi) + \phi
\end{equation}

The jet equation (\ref{jet}) is similar to a one dimensional equation
of motion for a particle ( with the position $\phi$ depending on a time $\tau$)  under the
action of a force $-dU/d\phi$ deriving from the potential,
\begin{equation}
\label{Uint}
U(\phi) = \frac{\ln(\cosh(C_0\phi))}{C_0} - \frac{\phi^2}{2},
\end{equation}
represented in figure \ref{conint}(b).
In its trajectory the particle energy is conserved :
\begin{equation}
\label{par}
\frac{1}{2}\left(\frac{d\phi}{d\tau }\right)^{2} + U(\phi) = Cst
\end{equation}
Let $\phi_i \equiv  \psi_i/R^2-B$, $i = -1,0,1$, corresponding to the solutions
$\psi_i$ of the algebraic equation (\ref{alg}). From (\ref{psiu}), we 
have $\phi_{\pm1} = \pm u$. Note that the stationary limit 
of (\ref{jet}),  which must be reached for
 $\lim_{\tau \rightarrow \pm \infty}$, yields again (\ref{uu}). Moreover, the 
particle energy conservation (\ref{par}) imposes the integrability condition,
\begin{equation}
\label{int}
U(\phi_{-1}) = U(\phi_1),
\end{equation}
which is indeed satisfied due to the symmetry of the potential $U$. We note
 that the Lagrange parameter determination (\ref{alpha0}) and the 
symmetry of the probabilities (\ref{p}) with respect to $\frac{1}{2}$
 could have been deduced from this integrability condition (\ref{int}) instead
 of minimizing the free energy (\ref{fp}) (we shall proceed in this way in section 3 to take into account
the beta-effect).

The jet equation (\ref{jet}) has been numerically integrated. Figure
\ref{figjet} shows a typical stream function and velocity profile in the
jets. Figure \ref{lar}(a) shows how the jet width depends on the segregation parameter $u$. We note that 
the width of the jet is an increasing function of the mixing  and
 therefore a decreasing function of the energy. Figure \ref{lar}(b) shows the dependence in $u$ of the total non-dimensional energy $e(u) = \frac{1}{2}\int_{- \infty}^{+ \infty} \left(d\phi/d\tau\right)^2 d\tau$ and of the maximum non dimensional jet velocity $\left(d\phi/d\tau\right)_{max}$. \\

As the jet structure (\ref{jet}) does not depend on the coordinate tangent to the jet, we
 can define the jet entropy ( respectively energy, free energy ) per unit
 length $S_{Jet}$ ( respectively $E_{Jet}$, $F_{Jet}$ ). Multiplied by the jet length, these
 quantities are the first order corrections to the entropy ( respectively energy, 
free energy ). Using
 the change of variables (\ref{var}), we calculate the jet entropy per  unit length :
$$
S_{Jet} = R \int_{- \infty}^{+ \infty}[s(p(\tau)) - s(p_{\pm 1})]d\tau 
$$
where $s$ is defined in (\ref{SA}), and $p_{\pm 1}$ are defined in (\ref{p}). Using the probability equation (\ref{probagibbs}) and (\ref{var})
we obtain :
\begin{equation}
\label{entropiejet}
S_{Jet} = R \int_{- \infty}^{+ \infty}[\tilde{s}(\phi)) - \tilde{s}(\phi_{\pm 1}))]d\tau
\end{equation}
involving the function $\tilde{s}(\phi) \equiv \ln(\cosh(C_0\phi)) - C_0\phi \tanh(C_0\phi)$.
Similarly we straightforwardly calculate the potential and kinetic energy per unit
length for the jet :
$$
E_{Jet}^P = \frac{R^3}{2}\int_{- \infty}^{+ \infty}(\phi^2 - \phi_1^2)d\tau \qquad 
E_{Jet}^K = \frac{R^3}{2}\int_{- \infty}^{+ \infty}\left(\frac{d\phi}{d\tau}\right)^2d\tau
$$
We use the integral (\ref{par}) to calculate $d\phi /d\tau$ and conclude :
\begin{equation}
E_{Jet} = R^3 \int_{- \infty}^{+ \infty}[\tilde{g}(\phi) - \tilde{g}(\phi_{\pm 1})]d\tau,
\label{energyjet}
\end{equation}
with $\tilde{g}(\phi) = - \frac{\ln(\cosh(C_0\phi)}{C_0} + \phi^2$.
Due to the symmetry of the jets, the jets provide no perturbation to the
 zero-order circulation, so there is no circulation term in the jet free
 energy expression : $F_{Jet} = -S_{Jet} - C_0/R^2 E_{Jet}$. Then 
\begin{equation}
F_{Jet} = C_0R \int_{- \infty}^{+ \infty} [\tilde{h}(\phi) - \tilde{h}(\phi_{\pm 1})]d\tau
\label{energylibre}
\end{equation}
with $\tilde{h}(\phi ) = -\phi(\phi - \tanh(C_0\phi))$.

Let us study the sign of $F_{Jet}$. As $\phi_{1}$ verifies $\phi_1 = \tanh(C_0\phi_1)$ 
and as $\phi (\tau)$
is an increasing function of $\tau$ with $\lim_{\tau \rightarrow + \infty}
\phi (\tau) = \phi_{1}$ we conclude that $\tilde{h}(\phi ) - \tilde{h}(\phi_{1}) > 0$ for
any $\tau > 0$. Thus $F_{Jet} >0$. Using the analogy with usual
 thermodynamics, the `surface tension' is positive. This favors large
 `bubbles' which
minimize the interfacial length and therefore 
the corresponding free energy (\ref{fp}). Our
 initial hypothesis of well separated domains with uniform $\psi$ is thus
 supported, as discussed more precisely in next subsection.

\subsubsection{Selection of the sub-domain shape}

\label{sel}

The above analysis has permitted us to determine the areas of subdomains on which
 the stream function $\psi$ takes the
constant values $\psi_{\pm 1}$, but the subdomains shape is
still to be selected. There is an analogy with two phases coexisting in thermodynamic 
equilibrium, for
instance a gas bubble in a liquid medium, for which a classical
thermodynamic argument explains the spherical shape of the
bubble by minimizing its free energy, proportional to the bubble area. Our
system is isolated rather than in a thermal bath, but the jet energy is small
(of order $R$) with respect to the total energy. Therefore the subdomain interior behaves
like a thermal bath with respect to the jet, so the usual argument on free energy 
minimization applies. We shall now show this more precisely by directly
maximizing the total entropy with
constraints, taking into account the jet contribution. 

The jet with length $L$
has an entropy $S_{jet}L$ and energy $E_{jet}L$. Since the total
energy  $E = E_{A}(C) + LE_{Jet}$ is given, the jet has also an indirect influence in the
 area energy
$E_{A}$. This small energy change $\delta E_A$ results in a corresponding change in the 
area entropy  $\delta S_A
= -({C_0}/{R^2}) \delta E_A$, from the condition (\ref{delF}) of zero 
first variations. Note that there is no area change $\delta A$ since the jet is symmetric and has
therefore no influence in 
the condition (\ref{PVcons}) of a given 
integral of $p$ (the difference in $p$ with the case of two uniform patches
with boundary at the jet center has zero integral). Therefore, adding the direct and indirect 
 contribution of the jet entropy leads to the total entropy
\begin{equation}
\label{S}
S = S_A(C_0) + \left(\frac{C_0}{R^2}E_{Jet} + S_{Jet}\right)L = S_A(C_0) - F_{Jet}L
\end{equation}
where $S_A(C_0)$ is the zero order, area entropy, obtained in the limit of vanishing jet width \footnote{ This reasoning to obtain the first order entropy can be precised by evaluating explicitly the first order modification of the Lagrange parameter $C$ ( let say  $C_1 \equiv C - C_0$ ) due to the jet energy, and the first order modification of the Lagrange parameter $\alpha$ ( let say  $\alpha_1 \equiv \alpha - \alpha_0$ ) due to the jet curvature and computing the first order entropy from its definition (\ref{ent}). We have calculated the first order entropy in this way to actually obtain (\ref{S}).}

We deduce from (\ref{S}) that the maximization of
 the entropy is achieved by minimizing the total free energy $F_{Jet}L$, which we
 have proved to be positive
at the end of previous sub-section. Thus we conclude that the maximum entropy state minimizes the
jet length, with a given area of the subdomains (\ref{aire}). The subdomains shape will 
therefore be a circle or a stripe.
More precisely if $A_{1} < 1/\pi$ the jet forms a circle enclosing the
positive constant stream function domain ( the jet bounds a cyclonic vortex ), if  $1/\pi < A_{1} < 1 -
1/\pi$ two straight lines jets form a stripe and if $A_{1} > 1 -
1/\pi$ the jet form a circle enclosing the
negative constant stream function domain ( the jet bounds anti-cyclonic vortex ).

The different types of solutions can be summarized in a (E,B) diagram : figure
\ref{phase}. The
 outer parabola is the maximum energy achievable for a
fixed B : $ E ={R^2}(1 - B^2)/2$. The frontier lines between the straight jets and the
circular jets corresponds to $A_{1} = 1/ \pi$ or $A_{-1} = 1/ \pi$. It
has been calculated using (\ref{aire}) and (\ref{eneu}) : $E =
R^2B^2({2\pi-2})/{(\pi -2)^2}$. Note that the maximum accessible energy is in $R^2$, but it has 
been scaled by the normalization
condition (\ref{csttemps}) on PV levels, so the real energy is not bounded.

All this analysis assumes that the vortex size is much larger than the jet 
width $\lambda$, given by figure \ref{lar}. In
 other words, the area $A_{1}$ or
  $A_{-1}$ (\ref{aire}) must be larger than $(2\lambda )^2$. This is satisfied
 on the right side of the
dashed line represented in figure \ref{phase}. The dashed line itself corresponds to
 the equality, and the
 condition
of large vortex is clearly not satisfied on its left side, for low energy. The position
 of the dashed
  line depends on the numerical value of $R$ (it has
been here numerically
  computed for $R = 0.03$), and it gets closer to the origin as $R\to 0$. We shall now determine 
the statistical equilibrium in this case of low energy.

\subsubsection{Axisymmetric vortices}

\label{secaxi}

 We have noted in subsection (\ref{secsub}) that, when for
instance $B>0$, in the limit of small energy ($E \rightarrow 0$ or equivalently
 $u \rightarrow |B|$, for fixed $B$ and $R$), the area $A_{-1}$ occupied by  $\psi_{-1}$
 tends to 1, the
whole domain. Therefore, in this limit, the complementary area $A_{1}$ tends to 0 and the
vortex becomes smaller than the deformation radius, so we can no more neglect the curvature
 radius of the
jet. 

 In this limit $u \rightarrow |B|$, as the vortex has a small area with
respect to the total domain, it is not affected by the boundary conditions, so it can be supposed
axisymmetric. From the general Gibbs states equation (\ref{gibbs}), we deduce the axisymmetric
vortex equation : 
\begin{equation}
\label{axi1}
- \frac{d^2\psi}{d\zeta ^2} - \frac{1}{\zeta} \frac{d\psi}{d\zeta } + \frac{\psi}{R^2} = B \ - \ \tanh \left(\alpha - \frac{C\psi}{R^2}\right)
\end{equation}
As $R$ will be a typical scale of the vortex, we make the
change of variable,
\begin{equation}
\label{varaxi}
\zeta = Rr \ ; \ \phi = - \frac{\alpha}{C} + \frac{\psi}{R^2}, 
\end{equation}
leading to the rescaled axisymmetric vortex equation :
\begin{equation}
\label{axi}
- \frac{d^2\phi}{dr^2} - \frac{1}{r} \frac{d\phi}{dr } + \phi + \frac{\alpha}{C}
= B + \tanh(C\phi))
\end{equation}
From now on, we shall consider the case
$B>0$ (the case $B<0$ is just the symmetric case of a negative vortex). 

For this equation
 to describe a localized
vortex, we impose $lim_{r \rightarrow \infty} \phi (r) 
= \phi_{-1}\equiv-\alpha/C+\psi_{-1}/R^2$, where $\psi_{-1}$ is the positive solution of the
 algebraic equation (\ref{alg}). Since nearly the whole fluid domain is
 covered by the asymptotic stream function
$\psi_{-1}$ outside the vortex, the condition of zero total circulation $\langle q \rangle =
\langle \psi \rangle / R^2 = 0$ imposes that $\psi_{-1}\simeq 0$ (it is of order $R$), so that
$\phi_{-1} = - \alpha/C$, and the algebraic equation (\ref{alg}) then leads to:
\begin{equation}
\label{alpha}
\alpha= \arg \tanh (B)
\end{equation}

We can thus eliminate $\alpha$ in (\ref{axi}), leading to an equation depending
 on two parameters, $B$ and $C$,
 \begin{equation}\begin{split}
\label{axi2}
 \frac{d^2\phi}{dr^2} =- \frac{1}{r} \frac{d\phi}{dr } -\tanh(C\phi)+ \phi-B + \frac{\arg \tanh B}
{C} 
\\\frac{d\phi}{dr}(r=0) = 0 \ and \ \lim_{r \rightarrow \infty} \phi (r)
= - \frac{\arg \tanh B}{C} \end{split}\end{equation}
where the regularity condition at $r=0$ has been included.
Let us consider, as in section (\ref{jetequation}), the analogy of
equation (\ref{axi2}) with a one particle motion with 'position' 
$\phi$ and 'time' $r$. 

The last four terms on the right-hand side of (\ref{axi}) can be written 
as the derivative
$-dU/d\phi$ of the potential,\begin{equation}
\label{Uaxi}
U(\phi) = \frac{\ln(\cosh(C\phi))}{C} - \frac{\phi^2}{2} 
+ \left(B-\frac{\arg \tanh B}{C}\right)\phi,\end{equation}
(represented in figure \ref{intaxi}), while the first term can be interpreted as
 a friction effect. Indeed, 
an integration of (\ref{axi2}) leads to:\begin{equation}
\label{potaxi}
U(\phi_{-1}) - U(\phi(r = 0)) = - \int_0^{+ \infty} \frac{1}{r}
\left(\frac{d\phi}{dr}\right)^2 dr < 0
\end{equation}
Thus, in figure \ref{intaxi}(a), the hatched area on the right side must be greater than the 
one on the left (since $(U(\phi_{1}) - U(\phi_{-1}))> U(\phi(r = 0)-U(\phi_{-1})>0$).
It is clear from figure \ref{intaxi} that this is possible only if $\phi_0<0$ and $\alpha/C<B$, 
or, using (\ref{alpha}),
$C>{\alpha}/{B} = {\arg \tanh B}/{B}$. The value $C={\alpha}/{B}$ corresponds
 to the integrability condition (\ref{int}) when the effect of jet curvature  
is neglected. This effect is now taken into account by the departure of $C$ from 
this value, which we shall denote $\Delta C\equiv C - {\arg \tanh B }/{B}$. Then 
$\Delta C > 0$ and we expect to recover the results of section 2.3.2 in the 
limit $\Delta C \rightarrow 0$. Moreover, we must reach a uniform stream function
at large distance, solution of the algebraic equation(\ref{alg}), so it must
 have three solutions. We see 
in figure \ref{intaxi} that the corresponding $\Delta C$ must not exceed a maximal value, denoted
$\Delta C_{max}$ . 

We can prove that for any $B>0$ and $ \arg \tanh B/B
 < C < \arg \tanh B/B + \Delta C_{max}$, equation (\ref{axi2}) has a unique solution. 
Such solutions have been numerically obtained for $B=0.75$ and
$0<\Delta C<\Delta C_{max}$. Corresponding stream function profiles are shown
in figure \ref{figaxi}. 

As $\Delta C$ is
 decreased from $\Delta C_{max}$ to zero, two stages can be seen in figure \ref{figaxi}.
 First the
 maximum value for the stream function is increased while the mean width of the vortex
 remains of the order of $R$. In a second stage, when
$\Delta C$ goes to zero, as we are closer to the integrability condition for big 
vortices (\ref{int}),
$\phi$ remains longer in the vicinity of $\phi_1$ so the vortex size
 increases. Note that the energy monotically increases as $\Delta C$ is decreased, first by an increase in 
the vortex maximum stream function and then by an increase in size. Finally
 the case of a jet with negligible curvature studied
 in subsection 2.3.2 is reached when $\Delta C \to 0$. 

In conclusion, we have shown that in the limit of small energy, with fixed $B$ and $R$,
 the Gibbs states are approximated by axisymmetric
 vortices, whose radial structure depends on the parameter $\Delta C$, which
  monotonically decreases from $\Delta C_{max}$ to 0 as energy is increased. 
 
\subsubsection{The linear approximation for the Gibbs states}

The previous discussion of axisymmetric vortices was concerned with the limit of small energy with
fixed $B$ and (small) fixed $R$. We consider now the limit of small $E$ and $B$, i.e. the neighborhood of the 
origin in the phase diagram of figure \ref{phase}. Then from (\ref{eneu}), $u \rightarrow |B| << 1$. Figure
 \ref{lar} shows that for $|u| << 1$, the jet width diverges and therefore
the jet tends to develop on
the scale of the whole domain, so the approximation of a localized jet, or an isolated 
axisymmetric vortex, falls
down. 

In this limit of small $E$ and $B$, we can however linearize the Gibbs
states equation (\ref{gibbs}), following the work of Chavanis and
Sommeria (1996) for Euler equation. After
linearization, solutions are expressed in terms
of the eigenmodes of the Laplacian, and only the first eigenmodes can be entropy maxima.

 These
results are unchanged by the linear deformation term $\psi/R^2$, so the work of Chavanis
and Sommeria (1996) directly applies here. With the periodic boundary conditions, the first eigenmode
 of the Laplacian, a sine function of one of the coordinates, for instance $y$, is
 thus selected. This corresponds to the low energy limit of the two jet configuration
 shown in figure
\ref{phase}. The next eigenmode, in $\sin(\pi x)\sin(\pi y)$, has the topology of the vortex
 states. A competition
between these two modes is expected in the neighborhood of the origin for small $E$ and $B$. Note
that the range of validity of the linear approximation is limited to a smaller range of 
parameters than 
in the Euler case, and this range of validity gets smaller and smaller as $R\to0$. The dominant solution
with uniform subdomains and interfacial jets relies by contrast on the tanh like relation between PV and 
stream function, and it is genuinely non-linear.

\section{The channel case}
We now consider the channel geometry, which represents a zonal band around a given latitude. It
 is then natural to introduce a
beta-effect, or a mean sublayer zonal flow ( topography ), with the term $h(y)$ in (\ref{QG}). We
 shall study the two cases of a linear $h(y)$ ( beta effect and/or uniform velocity
 for the sublayer flow ) or a quadratic $h(y)$. We shall follow the presentation for
 the periodic boundary conditions, stressing only the new features.

\label{cha}

\subsection{The dynamical system}

Let us consider the barotropic QG equations (\ref{QG},
\ref{dir} and \ref{u}) in a channel $D = [-\frac{1}{2},\frac{1}{2}]^{2}$ with
 the velocity ${\bf v}$ tangent to the boundary for $y = \pm \frac{1}{2}$ and
 1-periodicity in the zonal direction. Thus we choose for the boundary conditions 
a constant $\psi$, denoted $\psi_{b}$, the same on the two boundaries 
 $y = \pm \frac{1}{2}$.  We note that, due to these conditions, the
 physical momentum (\ref{vitessemoy}) is equal to zero. It is always
 possible to satisfy this condition by a change of reference frame with
 a zonal velocity $V$ such that it moves with the center of mass of the 
fluid layer, and a corresponding change of the deep flow, resulting in an
 additional beta effect $h \rightarrow h + \frac{Vy}{R^2}$.

As in section (\ref{per}), we need to specify the gage constant in the stream
function $\psi$, and we generalize the integral condition (\ref{psi0}) as,
\begin{equation}\label{mascha} \frac{\langle \psi \rangle}{R^2} - \langle h(y) \rangle = 0.
\end{equation}
The total mass $\langle \psi \rangle$ is then constant in time (but not the boundary value $\psi_b$
in general). With these
conditions, the Dirichlet
problem (\ref{dir}) has a unique solution $\psi$ for a given PV field $q$.  We note that the scale unit is chosen such that the area of $D$ is equal to 1.

The integral of any functions of the potential vorticity (\ref{cas}) is still
conserved. Let in particular $\Gamma$ be the global PV, or circulation :
\begin{equation}
\label{circ}
\Gamma \equiv {\langle q \rangle} = \int_D - \Delta\psi\; d^{2}{\bf r} = \int_{\partial D} {\bf v.}d{\bf l}
\end{equation}
By contrast with the doubly periodic boundary conditions, the circulation $\Gamma $ is not necessarily equal to 
zero. The expression of the energy in terms of the PV (see equation (\ref{ene})) is
 therefore modified (due to the boundary term in the integration by parts) :
\begin{equation}
\label{enecha}
E = {\frac{1}{2}}\int_D
\left[\ ({\bf \nabla}\psi)^2 + \frac{\psi ^2}{R^2} \ \right] d^{2}{\bf r} = {\frac{1}{2}}\int_D (q+h(y))\psi d^{2}{\bf r} - \frac{1}{2}\Gamma \psi_{b} \end{equation}  
Due to the invariance under zonal translation of the system, another conserved quantity exists :
\begin{equation}
\label{mom}
M = \int_D yq d^{2}{\bf r} 
\end{equation}
This constant moment fixes the `center of mass' latitude for the PV field.

\subsection{General form of the Gibbs states}

\label{sgich}

et us consider the statistical mechanics on a two PV level
configuration : the initial
states is made of patches with two levels of potential vorticity, $q =
a_1$ and $q =a_{-1}$, occupying respectively the areas $A$ and $(1 -
A)$ in $D$. We keep the normalization (\ref{csttemps}) and definition (\ref{B}) for $B$.
 Now, since the circulation $\Gamma$ is non-zero, the area $A$ is related to $B$ by
 $A=(1-B)/2+\Gamma/2$. The boundary term in the expression of
the energy (\ref{enecha}) leads to  an obvious change in the energy
variation (\ref{deltaE}). Let $\gamma$ be the Lagrange multiplier associated with 
the conservation  (\ref{mom}) of the momentum $M$.

Adapting the periodic case computations, we then calculate the probability equation and the Gibbs state equation : 
 \begin{equation}
 p = \frac{1-\tanh \left(\alpha' - \frac{C\psi}{R^2}  + \gamma y \right)}{2}
\label{probagibbscha}
\end{equation}
\begin{equation}
q=-\Delta \psi +\frac{\psi}{R^2} - h(y) = B \ - \ \tanh \left( \alpha' - 
\frac{C \psi}{R^2} + \gamma y \right)
\label{gibbscha}
\end{equation} 
with $\alpha'\equiv \alpha+C\psi_b/R^2$. These results generalize (\ref{probagibbs})
 and (\ref{gibbs}) of section 2.

 In the case of a Gibbs state depending on $x$, the Lagrange parameter $\gamma$ is related
 to a zonal propagation of the equilibrium
 structure. The statistical theory only predicts a set of equilibria shifted in $x$, but introducing 
the result back in the dynamical equation yields the propagation. Indeed the Gibbs state
equation (\ref{gibbscha}) is of the form 
$q = f(\psi,y)$, which can be inverted (as it is monotonous in $\psi$, see Robert 
and Sommeria (1991)) to yield:\begin{equation}
\label{psiq}
\psi = g(q) + R^2\frac{\gamma y}{C}\end{equation}where $g$ is a function of the potential
 vorticity. From this relation we calculate the velocity using (\ref{u}) : ${\bf v}
= R^2\frac{\gamma}{C}\hat{{\bf x}} - g'(q)\hat{{\bf z}} \wedge {\bf
  \nabla }q$. As PV is advected ( equation (\ref{QG}) ) we obtain :
\begin{equation}\label{vstructure}\frac{\partial q}{\partial t} + R^2\frac{\gamma}{C} \frac{\partial
  q}{\partial x} = 0
\end{equation}
Thus the PV field is invariant in a frame propagating with the zonal
 speed $V_{sr} = R^2\frac{\gamma}{C}$.

\subsection{The limit of small Rossby deformation radius.}

\label{smRos} 

In this sub-section we
 propose to analyze the Gibbs state equation (\ref{gibbscha}) in the limit of small deformation
 radius  ( $R<<1$ ). The main difference with the periodic case resides in the latudinally depending 
topography $h(y)$, resulting in two effects. Firstly the subdomains of uniform PV are
no more strictly uniform, and contain 
a weak zonal flow. Secondly the jet curvature is no longer constant in general, but depends 
on the local topography. Using the same boundary layer approximation as in the periodic
case, the Laplacian term in the Gibbs state equation (\ref{gibbscha}) will be 
neglected, except possibly
 in an interior jet and in the vicinity of the boundaries
  $y = \pm \frac{1}{2}$ ( boundary jets ). 

Outside such jets, (\ref{gibbscha}) reduces to an algebraic equation, like 
previously :\begin{equation}\label{psiy}\frac{\psi}{R^2} - h(y) = B \ - \ \tanh
  \left(\alpha^{\prime} - C\frac{\psi}{R^2} + \gamma y\right).\end{equation}
 This is like (\ref{alg}), replacing
 the constants
$B$ by $B+h(y)$ and $\alpha$ by $\alpha'+\gamma y$. The three solutions can be still visualized
by figure \ref{intaxi}, but the position of the straight line with respect to the tanh curve
now depends on $y$, due to the terms $h(y)$ and 
$\gamma y$. We assume that this dependence is linear in $y$ or varies on scales much larger 
than $R$ so that the Laplacian term remains indeed negligible.
The zero order Lagrange parameters $\alpha'$, $C$, $\gamma$, involved in this expression,
 can be obtained by directly maximizing the entropy by the same method as in
 section (\ref{secsub}). A relation between the jet curvature and topography is then 
obtained at first order. This approach is developed in 
appendix (\ref{amax}). 

However it is more simple to proceed differently : we start from the jet equation and show 
that its integrability condition provides the relation between the jet curvature and 
 topography. To catch this effect, we take into account  
the radius of curvature of the jet, denoted $r$, like in section 2.3.4, but state that $r$
is constant across the jet, assumed thin. From the Gibbs state equation 
(\ref{gibbscha}), using
 the boundary layer approximation, we thus obtain the jet equation in term of the transverse
coordinate $\zeta$ :\begin{equation}
\label{jetcha}
- \frac{d^2\psi}{d\zeta ^2} - \epsilon\frac{1}{r}\frac{d\psi}{d\zeta} + \frac{\psi}{R^2} - h(y) = B \ - \ \tanh
  \left(\alpha^{\prime} - C\frac{\psi}{R^2} + \gamma y\right).
\end{equation} We have introduced $\epsilon=\pm 1$ to account for the direction 
of curvature (keeping $r>0$). We define
$\epsilon=1$ (respectively -1) if the curvature of the jet is such that $\phi_1$ 
( respectively $\phi_{-1}$ )
is in the inner part of the jet. Note that, in the case of a vortex, as in our notations 
$\psi$ is proportional to the opposite of the pressure the case $\epsilon = 1$ ( resp
 $\epsilon \ = -1$ ) corresponds to a cyclone ( resp an anticyclone ). 

 The algebraic equation (\ref{psiy}) depends on three Lagrange parameters, instead of two 
for the periodic case of previous section, but we have three additional constraints, the
 condition that $\psi$ has the same
value at the two boundaries, the
 circulation constraint (\ref{circ}) and the momentum constraint (\ref{mom}). This will be
achieved in general
by boundary jets. Let us first study the interior jets.
 
 To study the interior jet, we make the change of variables
 :\begin{equation}
\label{varcha}
\tau \equiv \frac{\zeta}{R} \ ; \ \phi \equiv - \frac{\alpha^{\prime} + \gamma y }{C} + \frac{\psi}{R^2}
\end{equation}
We assume the variations of $y$ in the jet width are negligible ( $R <<$ scale of variation 
of $h(y)$ ), so that $y$ is treated as a constant. 
Then we obtain the jet equation :
\begin{equation}
\label{jet2cha}
- \frac{d^2\phi}{d\tau ^2} - \epsilon \frac{R}{r}\frac{d\phi}{d\tau} + \phi + \frac{\alpha^{\prime} + \gamma y}{C}  = B + h(y) + \tanh(C \phi)
\end{equation}
with $\phi \to \phi_{\pm1}$ for $\tau \rightarrow \pm \infty$, where again $\phi_{\pm1}$
 corresponds to the solutions of the algebraic equation (\ref{psiy}), rescaled as
\begin{equation}
\label{alg2cha} \phi + \frac{\alpha^{\prime}}{C}- B+\frac{\gamma y}{C}-h(y)= \tanh(C \phi).
\end{equation}

Let us consider, as in section (\ref{secaxi}), the analogy of the equation (\ref{jet2cha}) with
 the motion equation of a particle in the potential :
\begin{equation}
\label{potcha}
U(\phi) =  \frac{\ln \cosh(C \phi)}{C}-\frac{\phi^2}{2} + \left(B+ h(y) - \frac{\alpha^{\prime} 
 + \gamma y}{C}\right)\phi.
\end{equation} Like in section 2.3.4., integration of (\ref{jet2cha}) from $-\infty$ to $+\infty$
imposes the integrability condition :\begin{equation}\label{potrayon}U(\phi_1) - U(\phi_{-1})
 = \epsilon \frac{R}{r}\int_{-\infty}^{+\infty} \left(\frac{d\phi}{d\tau}\right)^2 
d\tau\end{equation}
 The second term of the l.h.s. of equation
 (\ref{jet2cha}) can be interpreted as a 
friction term: if $\epsilon=1$, the 'particle' starting from rest at $\phi_1$ can reach a state
of rest at $\phi_{-1}$ only if the difference of 'potential' corresponds to the 
energy loss (\ref{potrayon}) by friction (if $\epsilon=-1$ the same is true in
 the reversed direction).

  As in the periodic case we have made the thin jet assumption $R << r$, so
 that the friction term ( rhs of (\ref{potrayon})) is a correction of order $R/r$ : 
$U(\phi_1)-U(\phi_{-1}) = O(R/r)$. We first neglect it to get the zero order results, so we
write $U(\phi)=U(\phi_{-1})$. Therefore the two hatched areas in figure \ref{intaxi} must
 be equal, like in figure \ref{conint}. Due to the symmetry of the tanh function, this clearly
implies that the central solution of the rescaled algebraic equation (\ref{alg2cha}) must
 be $\phi_0=0$, so that $\alpha'_0/C_0-B+\gamma y/C_0 -h(y)=0$ (denoting the zero order Lagrange
parameters by the index 0). This is possible at different 
latitudes $y$ only if $\gamma y/C_0 -h(y)=0$, or is of order $R$ (so that it can be neglected at 
zero order). Then the integrability condition becomes
\begin{equation}\label{intcha}
\alpha_0^{\prime} = C_0B. 
\end{equation}
Furthermore $\phi_{\pm 1}$ are symmetric with respect to 0, of the form
 $\phi_{\pm 1} = \pm u$, determined by equation (\ref{uu}), like in section 2.3. This parameter
$u$ is again related to the energy by (\ref{eneu}). Finally, the terms
$\gamma y/C_0 -h(y)=0$ and the curvature term disappears in
 the jet equation (\ref{jet2cha}), which therefore reduces to (\ref{jet}), discussed in section 
 \ref{jetequation}. 

The first order solution outside the jet is obtained as a small correction  $\delta \phi_{\pm 1}$ 
to the zero
order solutions $\pm u$, with also a small correction $\alpha_1$ and $C_1$ to the parameters
 $\alpha'_0/C_0$ and $C_0$,
\begin{equation}
\label{firstor}
 \phi_{\pm 1} = \pm u + \delta \phi_{\pm 1}(y)\ ,\ \frac{\alpha'}{C} = \frac{\alpha_0'}{C_0} + \alpha_1\ ,\ C=C_0+C_1
\end{equation}
From (\ref{potrayon}), we deduce that $U(\phi_1)-U(\phi_{-1})$ has the
 sign of $\epsilon$. As may be seen in figure \ref{conint}, when $\alpha_1$ is positive, the line $\phi + \frac{\alpha^{\prime}}{C}$ moves upward, so that $U(\phi_1) < U(\phi_{-1})$. Thus
 $\alpha_1$ has the sign opposite to the sign of $U(\phi_1) - U(\phi_{-1})$ ; and
 we conclude that $-\epsilon \alpha_1$ is always positive.
Introducing this expansion (\ref{firstor}) in the algebraic equation (\ref{alg2cha}), using the zero
order results (\ref{uu}) and
(\ref{intcha}), we obtain : 
\begin{equation}
\label{deltaphi}
C_0\delta \phi_{\pm 1}(y) = \frac{-\alpha_1C_0 +C_1[B\pm C_0 u (1-u^2)]- \gamma y +
 C_0h(y)}{1-C_0(1-u^2)}
\end{equation}
Coming back to the stream function $\psi$, using (\ref{varcha}), we deduce the corresponding velocity ${\bf v}$ by differentiation with respect to $y$. This velocity outside the jet
is zonal, along the unit vector $\hat{\bf x}$, and verifies:
\begin{equation}
\label{shear}
{\bf v} = R^2\left(\frac{\frac{dh(y)}{dy} - \gamma (1-u^2)}{1-C_0(1-u^2)}\right)\hat{\bf x}
\end{equation}
It is therefore a constant plus a term proportional to the local beta-effect $\frac{dh}{dy}$. Notice that the corresponding shear $\frac{d{\bf v}}{dy}$ is stronger than the deep shear $d^
 \psi_d/dy^2 = R^2d^2h/dy^2$ by the factor $[1-C_0(1-u^2)]^{-1} > 1$.
The integrability condition (\ref{potrayon}) now provides the curvature of the 
jet. We can approximate the r.h.s. of this relation, of order $R$, by the zero order jet
 profile (\ref{jet}), denoting :
\begin{equation}
\label{eC}
e(u) \equiv \frac{1}{2}\int_{-\infty}^{+\infty} \left(\frac{d\phi}{d\tau}\right)^2 
d\tau .
\end{equation} 
( see figure \ref{lar}(b) ).
The l.h.s. can be expanded, using (\ref{firstor}). We first expand the expression (\ref{potcha}) of the 
potential, $U(\phi)=U_0(\phi)+ C_1/C_0 \left[\phi \tanh(C_0 \phi)- \log \cosh(C_0\phi)/C_0 \right] + \left[ h(y)- \gamma y/C_0 - \alpha_1 \right] \phi$. We can approximate $\phi\simeq \pm u$
 in the correction terms, and expand $U_0(\phi)=U_0(\pm u)+dU_0/d\phi(\pm u)$. The zero order
 equilibrium
condition imposes that $dU_0/d\phi(\pm u)=0$, so that (\ref{potrayon}) becomes
\begin{equation}\label{ry}\epsilon u \left(h(y)-\frac{\gamma y}{C_0}-
\alpha_1 \right) = e(u)\frac{R}{r}\end{equation}
This equation (\ref{ry}),
 expresses the dependence in latitude $y$ of the curvature radius $r$ of the curve on which
 the jet is centered, thus defining the shape of 
the sub-domain interface as a function of the topography.

Without topography and for $\gamma=0$, we get a constant jet curvature. The same result was
 obtained in
sub-section 2.3.3 by a different argument of free energy minimization. The parameter
 $u$, related to the energy by (\ref{eneu}) and to $C_0$ by (\ref{uu}),
 quantifies the strength of the jet. By contrast, the vortex area is determined by
the constraint on PV patch area (parameter $B$), but it is also related to the jet
 curvature, proportional
by (\ref{ry}) to the small shift $\alpha_1$ in chemical potential and 
temperature. Likewise the 
 equilibrium temperature at a liquid-gas interface slightly depends on the bubble 
curvature, due to capillary effects.

 As explained in the 
end of section (\ref{sgich}) the parameter $\gamma$ is linked to the zonal propagation speed of
 the structure. The term $\gamma y$ in (\ref{ry}), combined with a usual beta effect (linear topography 
term $h(y)$), leads to an oscillation with latitude $y$ of the jet curvature $1/r$, i.e. a 
meandering jet. Another possibility is an exact compensation of the beta-effect by
 the $\gamma y$ term, leading to a propagating circular vortex, and 
the selection between these two alternatives is discussed in next sub-section. An oval
shaped, zonally elongated vortex, such as on Jupiter, is obtained when this compensation occurs, but
with an additional quadratic topography $h(y)$. Indeed, to get a zonally elongated vortex, supposed
latitudinally centered in zero, the radius of curvature  of the jet 
must decrease for $y>0$ and 
increase for $y<0$. As a consequence, we deduce from (\ref{ry}) that the topography must be 
extremal at the latitude on which the vortex is centered (it actually admits a maximum in 
the cyclonic case and a minimum in the anticyclonic case ). Moreover, we deduce from
(\ref{shry}) that the surrounding flow must have a zero velocity at the latitude on which the 
vortex is centered and that the shear is cyclonic when the vortex is a cyclone and anticyclonic
 when the vortex is an anticyclone. More generally, the curvature can be related to the
 zonal velocity
 outside the jet, eliminating the topography between (\ref{shry}) and (\ref{ry}),
\begin{equation}\label{shry}{\bf v} = R^2\left(\frac{\gamma}{C_0} +
 \frac{\epsilon e(u)}{u\left(1-C_0(1-u^2)\right)}\frac{d}{dy}\left(\frac{R}{r}
\right)\right)\hat{\bf x}\end{equation} 

 Let us recall our approximations. Writing down the jet
 equation
 (\ref{jetcha}), when making the boundary layer approximation, we have assumed $R<<r$. We also 
assumed that in the jet width the topography can be considered as a constant. If $1/\sqrt a$
 is a typical length scale for the topography variation this gives $aR^2 << 1$. Moreover we have assumed
that the effective topography effect $h(y)-\gamma y/C_0$ remains small along the jet.
 If $L_V$ denotes the jet extension ( for example a vortex latitudinal size ) this 
approximation is valid 
as long as $aL_V^2 << 1$.      

\subsubsection{Beta effect or linear topography.}

\label{lineaire}Let $h(y) \equiv -\beta y$ in the following of this section. 
$\beta$ may mimic the beta-effect or a uniform velocity in 
the sublayer ( but we will refer it as the beta-effect ). \\

A first class of equilibrium states corresponds to a single
 solution $\psi(y)$ of the algebraic equation
 (\ref{psiy}). This determines a smooth zonal flow, with possibly intense jets at the boundaries
$y=\pm \frac{1}{2}$. The solution depends on the unknown parameters $C$, $\alpha'$ and $\gamma$, which 
are indirectly
determined by the energy $E$, momentum $M$, and the condition $\langle \psi \rangle =0$. The limit of small energy
corresponds to $C \to \infty$, for which we can neglect the term $\psi/R^2$ on the left
 hand side of (\ref{psiy}), which then reduces to 
$\psi= R^2/C[\arg \tanh(\beta y-B)+\gamma y+\alpha']$. This corresponds indeed to 
arbitrarily small values of $\psi$ (small energy) as $C\to \infty$.

When the particular energy value $E=R^2\beta^2/24$ is reached, a uniform PV is 
possible, with $\psi/R^2=-\beta y$. Then PV mixing is complete, which clearly maximizes 
the mixing entropy. In this case, 
$\gamma=-C\beta$, so that $\gamma y$ cancels the term 
$C\psi/R^2$ in (\ref{psiy}). Physically, the uniform westward zonal velocity, 
\begin{equation}
\label{um}
v_m = -R^2\beta
\end{equation}  
tilts the free surface with uniform
 slope by the geostrophic balance, and the corresponding topographic beta-effect exactly balances
the imposed beta-effect.
 
For a still higher energy, a first possibility is that again
\begin{equation}
\label{gamma}
\gamma = -C\beta
\end{equation}
so that the beta effect exactly balanced by the $\gamma y$ term in the jet equation of previous
subsection. This cancellation is directly obtained in the general Gibbs state
 equations (\ref{probagibbscha}) and (\ref{gibbscha}). Indeed the modified stream function
$\psi'=\psi+R^2\beta y$ satisfies the same equations as in the doubly-periodic case. Therefore
in the limit of small $R$, the Gibbs states are made of subdomains with uniform
$\psi'$ (uniform PV), separated by straight 
zonal jets or circular vortices. 

However jets or a vortex persist in this sea of uniform PV due to the constraint of
 energy conservation. The vortex moves westward at the same velocity $v_m$, according 
to (\ref{vstructure}) , so they
are just entrained by the background flow, without relative propagation (this can be
 physically understood by the cancellation
of the beta-effect). 
 
The selection of the subdomain areas and PV values is given like in the periodic case of section 2, just replacing
 $\psi$ by $\psi'=\psi-\beta R^2y$. Therefore we get again probabilities $p_{\pm1}=(1 \pm u)$ in the 
two subdomains with respective areas $A_{\pm1}$ given by (\ref{aire}), and stream function, 
\begin{equation}
\label{psiucha}
\psi_{\pm1} = R^2(B \pm u) - R^2\beta y 
\end{equation}
From this relation, we can calculate the energy $E =
\frac{1}{2}(\psi_{-1}^2A_{-1} + \psi_{1}^2A_{1})/R^2$, so the energy condition (\ref{eneu}) then becomes
\begin{equation}
\label{Eucha}
E = \frac{R^2}{2}(u^2 - B^2) +
\frac{R^2\beta^2}{24}
\end{equation}
Therefore these solutions with canceled beta effect can be obtained only beyond a minimum
 energy ${R^2\beta^2}/{24}$, corresponding to the potential energy of the surface tilting
 associated with the drift velocity $v_m$. Then the excess energy will control the organization
in two uniform PV areas.

The shape of these subdomains can be obtained again by minimization of the jet
 free energy. However, unlike in the periodic case, jets
occur at the boundaries $y=\pm \frac{1}{2}$ as well as at subdomain interfaces. Indeed, such boundary 
jets are in 
general necessary to satisfy $\langle \bf{v} \rangle =0$, or equivalently 
that the 
stream function $\psi_b$ must be equal at the two boundaries $y=\pm \frac{1}{2}$.  In particular, the 
solutions (\ref{psiucha}) necessarily involve a stream function difference (or mass flux) $-R^2\beta$ associated with
the drift velocity $v_m$. This stream function difference must be compensated
 by boundary eastward jets with opposite total mass flux. We show in Appendix \ref{appC} that for 
two PV levels with similar initial areas
a single eastward jet, separating two regions of uniform PV and weak westward drift, is the selected
state (instead of two opposite jets in the periodic case). In the case of a strong PV level with a small initial area, the system organizes in a circular vortex like in the periodic case. In 
the limit
 $u \rightarrow |B|$, as
one of the areas $A_{\pm 1}$ goes to zero, the jet approximation falls 
down. The corresponding analysis of 
axisymmetric vortices and of
the linear approximation for the Gibbs states, as performed in section 2, is still valid here.

Up to now we have ignored the constraint of the momentum $M$ (\ref{mom}). This constraint imposes the latitude $y_0$ of the equilibrium structure ( a circular patch or a zonal band with uniform PV ). For instance in the case
$B>0$, for which $A_1 > A_{-1}$ (as seen from (\ref{aire})),
we define  $y_0 \equiv \int_{A_1} y  d^2{\bf r}/A_1$. Then 
\begin{equation}
M \equiv \int_D yq d^2{\bf r} = \int_{A_1} y(B+u) d^2{\bf r} +
\int_{A_{-1}} y(B-u) d^2{\bf r} = 2uy_0A_1
\end{equation}
We thus deduce the latitudinal position of the equilibrium structure :
\begin{equation}
y_0 = \frac{M}{2uA_1} 
\end{equation}

In the case of a single eastward jet, the subdomain position has been already fixed
 by the area ($y_0=A_1/2$). Then the only possibility to satisfy a moment $M$ different from
$uA_1^2$ is that the jet oscillates in latitude with some amplitude $\Lambda$ (then 
$M-uA_1^2\simeq u\Lambda^2$ ). This is 
possible  if $\gamma\ne -\beta C$ according to (\ref{ry}), which becomes  
\begin{equation}
\label{rybeta}
 \frac{1}{r} = b(y-y_0) 
\end{equation} 
where $b \equiv - (u(\gamma + \beta C))/(2Ce(u)R) < 0$ and
 $y_0 \equiv \alpha_1/b$. This equation clearly leads to a jet oscillating 
 around the 
mean latitude $y_0$ ( as the curvature $r$ is positive for $y<y_0$ and negative for $y>y_0$ ; recall that the curvature is by definition positive when positive PV is in the inner part of the jet ). Note 
that this oscillation propagates eastward at speed $R^2\frac{\gamma}{C}$ ( given by (\ref{vstructure}) ). Since $b<0$, $\frac{\gamma}{C} > - \beta $, this speed is eastward with respect to the background drift $v_m$ (\ref{um}).


\subsubsection{Quadratic sublayer
 topography}\label{squ} As explained in section (\ref{smRos}), in the limit of small
 Rossby deformation radius $R$, the Gibbs state equation has solutions consisting of 
a vortex bounded by a strong jet on the scale of $R$. This corresponds to the case of an initial patch with strong PV and small area ( the asymmetry parameter $B$ is sufficiently large ) with an energy sufficiently strong to get a structure of closed jet ( see figure \ref{phase}). In the presence of a moderate topography $h(y)$, this internal jet is no more circular but its radius of curvature $r << R$ depends on $y$ according to (\ref{ry}). We have seen in previous subsection that a linear topography $h(y) = -\beta y$ leads to jets oscillating ( or to circular jets when $\gamma = - C\beta $ ). We shall discuss here how a quadratic term in $h(y)$ modifies the shape of closed jets. We therefore assume a topography $h(y)$ of the form :
 \begin{equation}
\label{topqua}h(y) \equiv ay^2 + by
\end{equation}
This corresponds to a uniform deep zonal shear, with velocity $v_d = R^2 d(h-\beta y)/dy = 2aR^2y + b - \beta$. We focus our attention on vortex solutions, seeking close curves solutions of equation (\ref{ry}). The vortices will be
 typically oval shaped as the ones seen on Jupiter. We then study how this shape ( for 
instance the ratio of the great axis of the oval to the small one ) depends on the 
topography ( sublayer flow ) and on the jet parameters. 
Application of these results to Great Red Spot observations will be discussed in next section.

To make equation (\ref{ry}) more explicit, let $s$ be a curvilinear parameterization of our curve, ${\bf T}(s)$ the tangent unit vector to the curve and $\theta(s)$ the angular function of the curve defined by ${\bf T}(s) = (\cos\theta(s),\sin\theta(s))$ for any $s$. Then the radius of curvature $r$ of the curve is linked to $\theta(s)$ by $1/r = d\theta /ds$ and (\ref{ry}) yields the differential equations :
\begin{equation}
\begin{align}\begin{cases}\label{thy}
\frac{d\theta}{dS} &= - dY^2 + 1 \\
\frac{dY}{dS} &= \sin\theta(S) \\
\end{cases}\\
\label{thx}
\text{and} \ \ \ \ \ \ \  &\frac{dX}{dS} = \cos\theta(S)
\end{align}
\end{equation}
with $c'X = x$, $c'Y = y-y_0$, $c'S=s$ ; where $d = (e^2(u)R^2a)/(\epsilon \alpha_1^3u^2)$, $y_0 \equiv e(u)R(C_0b-\gamma)/(\epsilon \alpha_1C_0u)$. The space coordinates $X$, $Y$ and $S$ are here non dimensional and have been obtained by dividing the real coordinates by the scale $c' \equiv e(u)R/(\epsilon \alpha_1u)$. Note that as explained in section (\ref{smRos}) $\epsilon \alpha_1 > 0$, so that $c^{\prime} >0$.  We further assume that $a>0$, so that $d>0$.

We first note that the two variables $\theta$ and $Y$ are independent of $X$. We will therefore consider the system formed by the two first differential equations (\ref{thy}). It is easily verified that this system is Hamiltonian, with $\theta$ and $Y$ the two conjugate variables and
 \begin{equation}
\label{H}
H \equiv \cos\theta - d\frac{Y^3}{3} + Y\end{equation}the Hamiltonian. Thus $H$ is constant on the solution curves. We look for vortex solutions of our problem (\ref{thy} and \ref{thx}).  Thus we require $\theta$ to be a monotonic function of $S$. Moreover the  curves  must close, that is $X$ and $Y$ must be periodic. For symmetry reasons, it is easily verified that the solutions of (\ref{thx},\ref{thy}) with initial conditions $\theta(0) = \frac{\pi}{2}$, $Y(0) = 0$  ( $H = 0$ ) and some $X(0)$ are periodic. We prove in appendix (\ref{appD}) that these initial conditions are the only ones leading to closed curves. We also prove that the solutions of (\ref{thy} and \ref{thx}) when $d > d_{max} \equiv \frac{4}{9}$ does not define $\theta$ as a monotonic function of $S$. They contain double points and thus are not possible solution for our problem.  Once given these initial conditions, we can easily prove that the structure has both a zonal symmetry axis and a latitudinal one passing through $y_0$.\\

To study the shape of the jets, we numerically solve equations (\ref{thy},\ref{thx}) with initial conditions : $\theta(0) = \frac{\pi}{2}, \ y(0) = 0$. We obtain closed curves with oval shapes, as  shown in figure \ref{ellipse}. In figure \ref{dimellipse} we have represented the width, the length and the aspect ratio of these vortices versus the parameter $d$. When $d$ tends to $\frac{4}{9}$, the vortex width tends to a maximum value : $w_{max} = \frac{3}{2}$ whereas the length diverges. In this limit, the vortices are thus very elongated.

\section{Application to the Jupiter's
  Great Red Spot and Oval BC}\label{secGRS}

In previous sections, we have found maximum entropy states with the
following properties :
 \begin{itemize}\item The fluid domain is partitioned  in two subdomains with 
   weak velocity, separated by jets whose width scales as the Rossby
   deformation radius. A strong initial PV level occupying a small
   area mixes in a subdomain with the form of a vortex bounded by an
   annular jet. \item In the presence of a parabolic topography $h(y)$
   (due to a sublayer zonal flow with uniform shear), outside the jet,
   exists a zonal flow (\ref{shear}) with uniform shear. The velocity
   at the latitude of the vortex center vanishes in the reference
   frame of the vortex.
 \item The curvature of the jet is linked to the topography by
   (\ref{ry}). For the parabolic topography, solutions are oval shaped vortices, symmetric in latitude and longitude.

\end{itemize}

These properties of our solutions are the main qualitative properties
of the Jovian vortices. Moreover, would this description be correct,
it would predict that the topography has an extremum at the center of
the vortex.

Dowling and Ingersoll (1989) derive the bottom topography, using
the GRS and Oval BC velocity fields obtained from cloud motion. They
analyze the results in the frame of a 1-1/2 shallow water model (SW),
with an active shallow layer floating on a much deeper layer. This
deep layer is in steady zonal motion which acts like a topography
$h_2$. The SW topography $h_2$ is defined by $fv_d = - 1/R_l(\lambda)\partial (gh_2)/\partial \lambda$ where $v_d$ is the deep layer flow and $R_l(\lambda)$ is the latitudinal radius of curvature of Jupiter. Dowling and Ingersoll (1989) have deduced this SW topography $h_2$ by assuming the conservation of
the shallow water potential vorticity $(\omega+f)/h_1$ ( $h_1$ is the upper layer thickness ) along the
streamlines of the observed steady vortex flow. The vorticity $\omega$
is deduced from the measured velocity field, and the planetary
vorticity $f$ is known, so that the variation of $h$ along each
streamline is deduced. The pressure field is then obtained from the Bernoulli relation and the hydrostatic balance, leading to the field $h_2$. The result
depends on the radius of deformation of Rossby ( $R^* = gh_0/f_0$, where $h_0$ is the mean upper layer height and $f_0$ the mean Coriolis parameter ), a free parameter in this
analysis. Three test values have been chosen, $R_1^* = 1700, R_2^* =
2200$ and $R_3^* = 2600$ km for the GRS and $R_1^* = 1100, R_2^* = 1600$ and
$R_3^* = 2000$ km for the Oval BC ( we denote by a star superscript the physical parameters, to distinguish them from the non-dimensional quantities used earlier ).  The height $h_2$ under each vortex has
been found to depend only on latitude, and has been fitted as a
quartic the planetographic latitude $\lambda$:
\begin{equation}\label{SWtop}gh_2 = A_0 + A_1\lambda + A_2\lambda^2 + A_3\lambda^3 + 
A_4\lambda^4.
\end{equation} 
The values 
obtained for the coefficients $A_i$ in the vortex reference frame, for each of the 
vortices and for each of the values $R_1^*, R_2^*, R_3^*$ are reported in table 1 of  Dowling and Ingersoll (1989).
 
Our model is the QG limit of this shallow water
system. Starting from (SW) equations, we can derive the QG equations
(\ref{QG}) by assuming the geostrophic balance and weak free surface
deformation in comparison with the mean layer thickness $h_0$. The
validity of this QG approximation has been discussed by Dowling and Ingersoll (1989)
and was found reasonably good as a first approach, although not
accurate. We furthermore use the beta-plane approximation, linearizing
the planetary vorticity around a reference latitude $\lambda_0$ (
$\lambda_0$ is taken to be $-23^0$ for the GRS and $-33.5^0$ for
the Oval BC). Therefore we write $f=f_0+\beta y$, with $f_0 =2\Omega
sin\lambda_0$ and $\beta \equiv 2\Omega \cos
  \lambda_0/r_z(\lambda_0)$ ( $\Omega$ is the planetary angular
speed of rotation : $2\pi/\Omega = 9$ h 55 mn 29.7 s and
$r_z(\lambda_0)$ is the zonal planetary radius, which slightly depends
on the latitude $\lambda_0$, due to the ellipsoidal planetary shape,
see formula (4) of Dowling and Ingersoll (1989) ) .  We then
obtain the QG potential vorticity (\ref{dir}) with the QG topography
$h^*(y^*)$ linked to the SW topography (\ref{SWtop}) by:
\begin{equation}\label{QGtop}h^*(y^*) = \frac{gh_2}{f_0R^{*2}} + \beta
  y^*\end{equation} 
 \\

We have computed the QG topography (\ref{QGtop}) using 
results of Dowling and Ingersoll (1989) for the SW topography (\ref{SWtop}) for the three values of
the Rossby deformation radius $R_1^*, R_2^*$ and $R_3^*$, for the GRS and
for the Oval BC. The result in figure \ref{figtopdow} shows that,
for both the GRS and the Oval BC, the QG topography has an extremum at
a latitude which is nearly the center of the vortex. As far as we
know, this fact has not been noticed in the literature.  This result
is in agreement with the predictions of our model. We note moreover
that the two extrema of the topography are minima, thus our model
predicts anticyclonic shears around the GRS and the Oval BC, as
observed.  Figure \ref{QGtop} shows a comparison of the QG
topography derived from Dowling results with a quadratic
approximation, in the case $R^* = R_2^* = 2200$ km.  This shows that the
quadratic approximation $h^*(y^*) = a^*y^{*2}$ is a good approximation on the
latitudinal extension of the GRS. This also provides values of the
parameter $a^*$ (\ref{topqua}) : $a^* = 9.2 \ 10^{-13}$, $a^* = 7.2\ 
10^{-13}$ and $a^* = 6.4\ 10^{-13}$ km$^{-2}$ s$^{-1}$ for $R^* =
R_1^*, R_2^*$ and $R_3^*$ respectively.
 
Let us deduce the corresponding non-dimensional
parameters. First, the PV levels were normalized by (\ref{csttemps}),
so our time unit $T^{\star}$ will depend on the real PV level
difference : $(a_1^{\star}-a_{-1}^{\star})/2 \equiv
\frac{1}{T^{\star}}$. The other parameters are the Rossby deformation
radius $R^{\star}$, the segregation parameter $u$ and the topography
coefficient $a^{\star}$ (\ref{topqua}).

We will consider $R^{\star}$ as a free parameter and use the following
data from GRS observation : \begin{itemize}
\item \underline{The jet width $l^{\star}$.} Let us define the width
  of the jet $l^{\star}$ as the width on which the jet velocity is
  greater than one half of the maximum velocity. We use 
  velocity measurement within the GRS of Mitchell et al (1981). They have used small clouds as tracers to measure
  velocities, and have observed that the velocity is nearly tangential
  to ellipses. Using
  a grid of concentric ellipses of constant eccentricity, the
  velocities have been plotted with
  respect to the semi-major axe $\cal{A}$ of the ellipse on which the
  measurement point lies. Results have been fitted by a quartic in $\cal{A}$
  and may be seen in figure \ref{vitMit}. Using these results,
  assuming that the velocity profile in the jet is symmetric with
  respect to its maximum, we choose as jet width $l^{\star} = 5.6\ 
  10^3$ km. In our model, the normalized jet width
  $l=l^{\star}/R^{\star}$ can be computed from the jet equation
  (\ref{jet1}) as shown in figure \ref{lar}(a). This
  determines the parameter $u$ from the parameter $R^{\star}$. The
  corresponding theoretical jet velocity can be compared to observations in figure \ref{vitMit}, for $R^{\star}=2500$ km (the shape is not very sensitive to this parameter). The computed results for $u$  versus
$R^{\star}$ is shown in figure \ref{figuR}. 
  
\item \underline{The maximum jet velocity $v^{\star}_{max}$.} We will
  use the value $v^{\star}_{max} = 110$ ms$^{-1}$ ( Mitchell et al 1981).  Using (\ref{var}) and giving real dimension gives
  :
\begin{equation}\label{temps}v^{\star}_{max} =
\frac{R^{\star}}{T^{\star}}\frac{d\phi}
{d\tau}\arrowvert _{max}(u).\end{equation}
$d\phi/
d\tau \arrowvert _{max}(u)$ has been obtained by solving the non-dimensional jet equation (\ref{jet1}), and is shown in figure \ref{lar}(b). 
This now determine $T^{\star}$ from $R^{\star}$.

\item \underline{The velocity shear surrounding the vortex.} 
The ambient zonal shear measured at the latitude of the  GRS from Limaye et al. (1986) is $\sigma^{\star}= 1.5e-5$ s$^{-1}$. Using (\ref{shear}) in its
  dimensional form, for a quadratic topography gives
  :\begin{equation}\label{astar}a^{\star} = 
\frac{\sigma^{\star}}{2R^{\star}} \left(1-\frac{C_0}{\cosh ^2(C_0u)}\right)\end{equation}
This permits to compute $a^{\star}$ as a function of $R^{\star}$ (since $u$ has been determined, as well as $C_0$, related to $u$ by (\ref{uu}) ).
\end{itemize}
The computed results for $a^{\star}$ and $T^{\star}$ versus
$R^{\star}$ are shown in figure \ref{figaa1-R}.  This shows that our
determination of the topography is in agreement with the QG topography
deduced from the shallow water model of Dowling and Ingersoll (1989) within a factor of two.  The
corresponding PV level difference $a_1^{\star} - a_{-1}^{\star}$ is
comparable to the planetary vorticity $f_0$ at the latitude of the center of
the GRS when $R \simeq 2400$ km. For this value of $R^*$, figure \ref{figuR} shows that $u$ is very close to 1. Furthermore, as the GRS area is very small compared to the global area of a latitudinal band centered around the GRS, the non dimensional area occupied by the positive PV is very close to 1. Using this area expression (\ref{aire}) we conclude that $B$ is very close to 1. Using the definition of $B$ (\ref{B}), we conclude that  $a_1^{\star} \simeq 0$ and $a_{-1}^{\star} \simeq -f_0$. As discussed below a forcing
mechanism by convective plumes incoming from the sublayer is expected
to yield this result.\\

The shape of the jet depends on the parameter $d  = \frac{e^2(u)}{\epsilon \alpha_1^3u^2}\frac{a^{\star}R^{\star ^2}}{a_1^{\star} -a_{-1}^{\star}}$ and on the length scale $c^{'*} \equiv \frac{e(u)R^*}{\epsilon \alpha_1u}$. We can determine these two parameters from the previously determined values of $a^*$,$T^*$ and $u$, and from the observed half width of the Great Red Spot : $y^{\star}_{max} = 4900$km. This permits to calculate $c^{\prime *}$, $\alpha_1$ and $d$ versus $R^*$. Figure \ref{figdR} show 
$d$ versus $R^{\star}$. The dot line represents the critical value $d
= \frac{4}{9}$ beside which a vortex solution exists. The ratio of the
length to the width of the GRS is approximatively $2$, which would correspond to $d=0.441$ ( figure \ref{dimellipse} ) ; this is very close to the critical value $\frac{4}{9}$. From figure \ref{figdR}, our model predicts that the Rossby deformation radius
is $R^{\star} = 1800$ km. Figure \ref{ellipse} shows the actual shape of the vortex for $d=0.441$. 

However, in the jet shape analyze, to obtain (\ref{thy},\ref{thx}) we have supposed $a^{\star}L_V^{\star ^2} << ( a_1^{\star} - a_{-1}^{\star})/2$ where $L_V^{\star}$ is the
maximal latitudinal extension of the vortex ( the topography part of
the PV remains negligible with respect to the PV ). For the value of
$R^{\star}$ calculated above, we find $a^{\star}{y^{\star}_{max}}^2 =
3.9 10^{-5} s^{-1}$ whereas $a_1^* - a_{-1}^* = 1.3 10^{-4} s^{-1}$. We are
thus at the limit of validity of our assumption.

Now we can reverse the procedure and propose a predictive model of the
Great Red Spot. Assume a steady deep zonal flow with uniform shear,
$v_d=2a^*R^{*2}(y^*-y_I^*)$, vanishing at an origin $y_I$ depending on our reference frame (which we shall choose in order to cancel the vortex drift). This flow is presumably
generated by deep thermal convection  but we are concerned here only
with the dynamics of the upper layer, assumed stably stratified (due
to cooling by radiative effects). We model this stratified upper layer
as a shallow layer with radius of deformation $R^{\star} \simeq 2400$ km.
This layer is submitted to a total beta effect or 'topography'
$h(y)=ay^2-(\beta+2ay_I) y$.

Assume that PV spots with value $-f_0$, occupying small area proportion are randomly generated in this layer. This would be
the result of intense incoming thermal plumes,  as recently discussed by Ingersoll et al (2000), : conservation of the
absolute angular momentum during the radial expansion leads to strong
decrease of the local absolute vorticity, which comes close to zero.
This means that in the planetary reference frame, a local vorticity
patch with value $-f_0$ is created. The opposite vorticity is globally
created by the subducting flow, but it is close to 0 due the much larger area. This gives our time unit $(2f_0)^{-1}$
and $B= 1 -2A$.

The outcome of random PV mixing with the constraint of the
conservation laws is then a zonal velocity (\ref{shear}) in the observed upper
layer and a vortex with area (\ref{aire}) with velocity profile shown by the dot
curve in fig 14.  The vortex moves with the upper velocity at $y=0$, it drifts with respect to the deep layer at velocity (\ref{um})   so that the beta effect is suppressed. The
shape is an oval symmetric in $x$ and $y$, with aspect ratio computed from the shape parameter $d$ ( figure \ref{dimellipse} ).

Note that a slightly larger area , or stronger energy could lead to
very elongated vortices. Then argument of free energy minimization show that this would lead to a single eastward zonal jet.
This may explain the jet observed in the Northern hemisphere of
Jupiter  at the same latitude as the GRS. 

 For smaller Jovian vortices such as the Oval BC, the size of
the vortices is comparable with the Rossby deformation radius. Thus
such vortices may be described as done for axisymmetric vortices ( section
\ref{secaxi} ). This explain why such vortices do not have a quiescent core
as the GRS.  

 Let us describe dark brown cyclonic spots ( 'Barges' ) at $14^°$ N on
Jupiter. Their first interest is to stress that cyclonic vortices
embedded in cyclonic shear exist on Jupiter. Let us go further.  The
greater of these barges is studied from Voyager observations by Hatzes
at al (1981). 
The meridional velocities measured at the latitude of the center of
the barge ( Hatzes et al (1981), figure 7 ) show a boundary
jet organization around the perimeter of the barge ( $v_{max} = 25$
ms$^{-1}$ ), see figure \ref{jets}(b). The surrounding shear is such that the shear velocity at
the maximum latitude of the barge is the same as the maximum jet
velocity. Thus our approximation $aL_V^2 << 1$ is not good. We can however
explain the elongated shape, 
similar to figure \ref{ellipse}(b) obtained for $d$ very close to
$d_{max}$.\\

We conclude that the Gibbs state equation (\ref{gibbscha})
derived from maximization of entropy of the QG model (\ref{QG}) is in
the limit of small Rossby deformation radius a model that explains the
main qualitative features of Jovian vortices.  The statistical
mechanics itself explains the organization of a turbulent flow in
coherent structures.

\section{Conclusion}

Our first result is to provide a general explanation for the emergence
and robustness of intense jets in atmospheric or oceanic turbulent
flows. In the absence of topography or beta-effect turbulence mixes
potential vorticity in subdomains, and such jets occur at the
interface of these subdomains, with a width of the order of the
deformation radius. From a thermodynamic point of view, this is like
coexistence of two phases. Indeed the vortex interaction becomes short
ranged in the limit of small deformation radius, and statistical
mechanics leads to a thermodynamic equilibrium between two 'phases',
with different concentrations of the 2 Potential Vorticity levels. Another approach
leading to the same result is to consider the general partial
differential equation (\ref{gibbs})  characterizing the equilibrium states.
This equation reduces to the algebraic equation (\ref{alg}) in the limit of
small deformation radius. The two uniform subdomains correspond to two
solutions $\psi_{-1}$ and $\psi_1$ of this equation. At the interface
of these subdomains, the general pde reduces to the equation (\ref{jet}),
whose solution determines the jet profile. In addition, a solvability
condition of this equation confirms the relation of equilibrium
between the two 'phases', which was obtained in the thermodynamic approach.

All our results have been obtained for a 2 potential vorticity level case, but cases with
more levels would lead to qualitatively similar results, although the
quantitative analysis would be more involved due to the additional
parameters.

In the presence of beta effect or topography,  for low energy, purely zonal flows
with gentle variation in latitude are obtained. A critical energy is the energy of the state where the zonal flow just compensates the beta-effect. For this state the PV is strictly uniform in the whole domain. For greater energy, two well mixed domain separated by jets appears, as in the without topography case. However the PV is no more strictly uniform in the well mixed subdomain : a zonal flow exists. Also the jet
curvature depends on latitude. With ordinary beta effect this yields
an intense eastward jet, purely zonal or wavy depending on the
constraint on the momentum $M$. With the quadratic beta-effect
generated by a deep shear, this can produce an oval shaped vortex. This vortex then drift so as to compensate the beta-effect. In other words, in the vortex reference frame the equivalent topography $h(y)$ admits an extremum, and this is in agreement with the data of Dowling and Ingersoll (1989).

Our quasi analytical approach therefore explains most of the basic features of the Great Red Spot and other Jovian vortices. It can be developed into a more accurate predictive model along the following lines. First the approximation $R<<r$ of thin jet is convenient for a qualitative understanding  but is only marginally satisfied. However this limitation can be overcomed by numerical determination of the equilibrium state equation (\ref{gibbs}) by methods like used by Turkington and Whitaker (1995) or using relaxation equations toward the maximum entropy state as described by Robert et Sommeria (1992). Furthermore, extension to the more general shallow water model is desirable, as the Rossby number ( $\simeq 0.36$ where it a maximal ) is not very small. This can be formally achieved ( in preparation ).

Finally the results rely on an assumption of ergodicity, or complete potential vorticity mixing consistent with the constraints on the conservation laws. Various numerical and laboratory experiments in the case of Euler equations ( see e.g. Brands et al. 1989 ) indicate that mixing may not be global but be restricted to active regions. Organization into local vortices, rather that at the scale of the whole domain, is more likely with a small radius of deformation, as vortex interactions leading to coalescence are then screened. This is observed for instance in the numerical computations of Kukharkin Orszag and Yakhot (1995). By contrast, the zonal shear here promote vortex encounters ( as observed in Voyager data ) and we expect a much better relaxation toward the global statistical equilibrium, which involves always a single vortex in a given shear zone ( as it minimizes the interfacial free energy ).

 \section{Acknowledgments}

 The authors thank R. Robert for collaboration on statistical mechanics approach and for useful comments on the present work.

\begin{appendix}

\section{Determination of the Gibbs state by direct entropy maximization, 
in the presence of a topography ( beta-effect ) .}

\label{amax}

In section (\ref{smRos}), we studied the limit of
 small Rossby deformation radius in the Gibbs state equation (\ref{gibbscha}) by considering the jet equation (\ref{jetcha}) and its integrability condition (\ref{potrayon}). We deduced
 that the Gibbs states are composed of subdomains in which $\psi$ verifies the algebraic 
equation (\ref{psiy}) separated by an interfacial jet whose curvature verifies (\ref{ry}). 
The aim of this annex is to prove that these results can be obtained by directly maximizing
 the entropy, adapting the method used in section (\ref{secsub}).
\\Let us make the following assumptions :
\begin{enumerate}\item In the limit of small 
Rossby deformation radius, the probability $p$  of finding the PV level $a_1$ takes 
two values $p_{\pm 1}(y)$, depending only on $y$. We are looking for vortex solutions.
 The vortex shape is described by the length $l(y)$ on which the probability $p$ takes 
the value $p_{- 1}(y)$ ( See figure \ref{figly} ). \item The two subdomains where $p$
 take the two values $p_{\pm 1}(y)$ are separated by a jet. The probabilities $p_{\pm 1}(y)$
 are supposed to be close to the their values without topography  $p_{\pm 1} = \pm u$, such 
that the free energy per unit length of the jet is well approximated by the one calculated 
without topography (\ref{energylibre}). If $L_V$ denotes the vortex size, $1/\sqrt a$
 a typical length on which topography varies, we will show this approximation to be valid as
 soon as $aL_V^2 << 1$.\item The boundary conditions can be relaxed ; that is no boundary 
term appears in the variation of the free energy at the order considered here.
 ( See discussion concerning boundary jets in section (\ref{smRos}).
\\\end{enumerate}

Given these hypothesis, the Gibbs states is described
by the 3 functions $p_{\pm 1}(y)$ and $l(y)$. We will determine them by maximizing the
 entropy $S$ (\ref{ent}) under the 3 constraints : energy (\ref{enecha}), mass conservation
 (\ref{mascha}) and momentum (\ref{mom}). A necessary condition for a solution to this 
variational problem is the existence of 3 Lagrange parameters $C$,$\alpha$ and $\gamma$ 
such that the first variations of the free energy :\begin{equation}
\label{Fcha}
F \equiv -S - \frac{C_0}{R^2}E + \alpha_0 \frac{\langle \psi \rangle}{R^2} + \gamma M\end{equation}
vanish. Using (\ref{ent}),(\ref{enecha}),(\ref{mascha}) and (\ref{mom}) ; (\ref{dir}) and
 (\ref{PVmean}) and the above hypothesis ( see figure \ref{figly} ), a direct calculation
 shows that the free energy (\ref{Fcha}) is up to a constant :\begin{equation}
\begin{split}
\label{fpcha}
F &=  \int_{y_{min}}^{y_{max}} \left[ f(p_{1}(y),y)(1-l(y)) +
 f(p_{-1}(y),y)l(y) \right]dy + \int_{-\frac{1}{2}}^{y_{min}} f(p_{1}(y),y)dy + \\&
 \int_{y_{max}}^{\frac{1}{2}} f(p_{1}(y),y)dy + LF_{Jet}(u) \\ 
 \text{with} \ \  f(p,y) 
&\equiv (p \log p + (1-p)\log (1-p)) - 2C_0\left(p - \frac{1}{2}\right)^2 - 2\left(C_0B - 
\alpha_0\right) \left(p - \frac{1}{2}\right) - \\
& 2\left(C_0h(y) - \gamma y\right)\left(p - 
\frac{1}{2}\right)
\end{split}
\end{equation} 
and where $L$ is the jet length, $F_{Jet}(u)$
 is the jet free energy per unit length (\ref{energylibre}), calculated without topography.
\\Considering first variations of the free energy (\ref{fpcha}) under variations of $p_{1}(y)$
 ( resp $p_{-1}(y)$ ) proves that $\partial f/\partial p(p_1(y),y) = 0$ and that 
$\partial f/\partial p(p_{-1}(y),y) = 0$.  A 
direct calculation shows that :
\begin{equation}
\label{uucha}
2\left(p_{\pm1}-\frac{1}{2}\right) = \tanh \left(2C_0\left(p_{\pm1}-\frac{1}{2}\right) + C_0h(y) - 
\gamma y + C_0B - \alpha \right)
\end{equation}
 Using (\ref{PVmean}) and (\ref{dir}) ; recalling that we neglect the Laplacian
 term, a straight calculation shows that (\ref{uucha}) is equivalent to the algebraic
 equation (\ref{psiy}).\\

Let us consider now first variations of the free energy 
(\ref{fpcha}) under small variations $\delta l(y)$ of $l(y)$. Using that the length
 of the jet is given by $L = 2\int_{y_{min}}^{y_{max}} \sqrt{1+\frac{1}{4}
\left(dl/dy\right)^2}dy$, a straightforward calculation shows that
 $\delta L = - \int_{y_{min}}^{y_{max}}\delta l(y)/r dy$ where $r$ is
the radius of curvature of the jet. We thus deduce  from first variations of 
the free energy (\ref{fpcha}) : \begin{equation}\label{ry1}\frac{F_{Jet}(u)}{r}
 = f(p_{1}(y),y) - f(p_{-1}(y),y)\end{equation}
Hypothesis (2) : $aL_V^2 << 1$ 
permitted us to consider $F_{Jet}(u)$ as independent of $y$. In accordance with 
this hypothesis we evaluate $f(p_{1}(y),y) - f(p_{-1}(y),y)$ at order zero, with
 $p_{\pm 1} = \frac{1}{2}(1\pm u)$ at this order. We obtain : $f(p_{1}(y),y) - f(p_{-1}(y),y) = 2u(\alpha1 +
 C_0h(y) - \gamma y)$. Moreover using the free energy per unit length expression 
(\ref{energylibre}), (\ref{Uint}) and (\ref{par}), one can show that $F_{Jet}(u)=2e(u)C_0R$
 where $e(u)$ is defined by (\ref{eC}). These two last results show that (\ref{ry1}) is
 equivalent to (\ref{ry}) the expression for the radius of curvature $r$ found by the 
integrability condition for the jet.

\section{Boundary jets in the channel case. 
Beta-effect or linear sublayer topography.}
\label{appC}Let us derive the boundary jet properties in the case of a beta-effect (or
 linear sublayer topography). For the sake of simplicity, we will treat the case of zero
 circulation  $\Gamma = 0$. Let $\phi_{b}^+$ and $\phi_{b}^-$ be the values of $\phi$ on the 
$y=\frac{1}{2}$ and $y=-\frac{1}{2}$ boundary respectively.
The boundary jet satisfies the jet equation (\ref{jet}), but with
boundary  conditions $\phi(\tau = 0) = \phi_{b}^{\pm}$ and $\phi(\tau 
\to +\infty) = \phi_{1}\; or\; \phi_{-1}$. Thus, using (\ref{par}), we deduce that :
\begin{equation}
\label{enejetfrcha}
\frac{1}{2}\left(\frac{d\phi}{d\tau}\right)^{2}\left(y = \pm \frac{1}{2}\right) \  + \ U(\phi_{b}^{\pm}) = U(\phi_{1}) = U(\phi_{-1})
\end{equation} 
(where the last equality comes from the integrability condition for the interfacial jet).

We can relate $d\phi/d\tau$ to the derivative $d\psi/d\zeta$  normal
 to the boundary ( denoting the normal coordinate $\zeta = \pm\frac{1}{2} \mp y$ ), using 
(\ref{varcha}),
\begin{equation}
\label{derfrcha}
\frac{d \phi}{d \tau}\left(y = \pm\frac{1}{2}\right) = \frac{1}{R} \frac{d \psi}{d\zeta}\left(y 
= \pm \frac{1}{2}\right) \ \mp \ R{\gamma\over C} 
\end{equation}
Furthermore the condition $\Gamma=0$ imposes
$$
\Gamma = -\int_{y=\frac{1}{2}} \frac{d \psi}{dy}dx + \int_{y=-\frac{1}{2}} \frac{d \psi}{dy}dx 
 =  \frac{d \psi}{d\zeta}\left(y=\frac{1}{2}\right) + \frac{d \psi}{d\zeta}\left(y=-\frac{1}{2}\right) = 0,
$$
so that  
$\frac{d \psi}{d\zeta}(y=\frac{1}{2}) = - \frac{d \psi}{d\zeta}(y=-\frac{1}{2})$.
Then (\ref{enejetfrcha}) becomes :
\begin{equation}
\label{frdetcha}
U(\phi_{b}^{\pm}) = U(\phi_{1}) - \frac{1}{2}\left(\frac{1}{R}\frac{d \psi}{d\zeta}\left(y=\frac{1}{2}\right)
 - R{\gamma\over C} \right)^2
\end{equation}
so that
\begin{equation}
\label{uphifrcha}
U(\phi_{b}^{+}) = U(\phi_{b}^{-})
\end{equation}
Note that the two 
values $\phi_{b}^{\pm}$ cannot be equal. They must indeed satisfy, from (\ref{varcha}) and the 
condition of zero mass flux ($\psi^+=\psi-$),
\begin{equation}
\label{deltaphifr}
\phi_{b}^+ - \phi_{b}^- = -\gamma/C
\end{equation}

To solve these two equations (\ref{uphifrcha}) and (\ref{deltaphifr}), let us have a look at
 the potential $U$ in figure 19. We have to compare $\phi_{b}^+ - \phi_{b}^-$ 
and $\phi_{1} - \phi_{-1}$. We thus distinguish two cases. Using (\ref{deltaphifr})
 and $\phi_{\pm 1} = \pm u$, we calculate :
 $(\phi_{b}^+ - \phi_{b}^-)/(\phi_{1} - \phi_{-1}) = -\gamma/(2Cu)$.
 We recall that $u$ is of order unity. 
\begin{itemize}
\item{\underline{Low beta-effect case : $-\gamma/C < 2u$.}}\\
For an x-independent statistical equilibrium, one possibility is a zonal band with $\phi_{-1}$ inside the band and $\phi_{1}$ near both boundaries, with $\phi_{b1}^+$ and  $\phi_{b1}^-$ at the boundaries. The symmetric solution, with $\phi_{b-1}^+$ and  $\phi_{b-1}^-$ at the boundaries and $\phi_{-1}$ near the boundaries, has the same free energy ( due to the symmetry of the potential $U$ ). However the solution which maximizes the total free energy of the jets corresponds to $\phi_{-1}$ in the lower part of the domain, and $\phi_{1}$ in the upper part, with $\phi_{b0}^+$ at the upper boundary ( $y = \frac{1}{2}$ ) and  $\phi_{b0}^-$ at the lower boundary ( $y = -\frac{1}{2}$ ), and a single eastward interior jet.

\item{\underline{High beta-effect case : $\beta > 2u$}}\\ 
In this case, we note that  $\phi_{b}^+ - \phi_{b}^- > \phi_{1} - \phi_{-1}$. Then
 the two equations (\ref{deltaphifr}) and (\ref{uphifrcha}) determine a unique 
solution for $(\phi_{b}^+,\phi_{b}^-)$ ( see figure 19 ). We are then necessarily in the case with $\phi = \phi_1$ near the boundaries $y = \frac{1}{2}$ and 
 $\phi = \phi_{-1}$ near the other one. It again involves a single eastward internal jet.
This jet can oscillate in latitude due to the momentum constraint ( according to (\ref{rybeta}) ).
\end{itemize}

Once $\phi_{b}^{\pm}$ are fixed, we use equation (\ref{varcha}) on the two boundaries 
$y = \pm \frac{1}{2}$ to conclude that $\alpha = - \frac{1}{2}(\phi_{b}^+ + \phi_{b}^-)$.
 Then using the integrability condition (\ref{intcha}) we calculate : 
\begin{equation}
\psi_{b} = R^2\left( B + \frac{\phi_{b}^+ + \phi_{b}^-}{2} \right)
\end{equation}
This closes the determination of our parameters.\\

\section{Boundary condition for the equation (\ref{thy},\ref{thx}) of the oval-shaped vortices boundary}

  \label{appD}

We want to see whether equations (\ref{thy},\ref{thx}) with $d>0$, defining the curve formed by the jet, admit periodic solutions both in $x$ and $y$ corresponding to vortices, or not. As stressed in section (\ref{squ}), equations (\ref{thy}) derive from the Hamiltonian (\ref{H}) ; $y$ and $\theta$ are the two conjugated variables. Let us study the phase portrait of $H$. For $\theta$ in $[0,2\pi [$, there are 4 critical points : $P_1 = (0,1/d),\ P_2 = (0,-1/d),\ P_3 = (\pi,1/d),\ P_4 = (\pi,-1/d)$. By linearization around these fixed points, one easily prove that $P_1$ and $P_3$ are stable fixed points whereas $P_2$ and $P_4$ are hyperbolic fixed points. This permits to draw the phase portrait : figure \ref{Hphase}. Using $H$ (\ref{H}), we obtain that the unstable manifolds are given by $1-2/(3\sqrt{d}) = H$ and $-1+2/(3\sqrt{d}) = H$ respectively. The parameter $d$ governs a transition of the phase space structure. This transition occurs when the two unstable manifolds merge ; this permits to compute the transition value for $d$ : $d = \frac{4}{9}$. We are looking for vortices solutions of (\ref{thy},\ref{thx}). We recall that $\theta$ is the angle with the $x$ axe. We thus impose $\theta$ to be a monotonic function of $s$ on a trajectory. Thus areas $c$ of the phase portrait (figure \ref{Hphase} ) are forbidden ; they would correspond to $\theta$ varying in a finite interval. We won't address their analyze there, but oscillating jet solutions can be found in these areas. Area $b2$ : it can be shown that, as on such trajectories, as $\theta$ is not strictly increasing, double points exists, thus forbidding area $b2$. Conversely areas $a$ and $b1$ are admissible trajectories, giving $y$ as periodic functions of $\theta$.We have to had the condition that $x$ (\ref{thx}) must also be a periodic function of $\theta$ for the curve to close. Let us denote $\Delta x$ the $x$ variation when $\theta$ runs in $[0,2\pi ]$. We thus impose the condition $\Delta x = 0$. Using (\ref{thx}) and (\ref{thy}) we calculate : 
\begin{equation}
\label{Deltax}
\Delta x = \int_0^L \cos \theta ds = \int_0^{2\pi } \frac{\cos \theta d\theta}{-dy^2(\theta) + 1} = \int_{-\frac{\pi}{2}}^{\frac{\pi}{2}} \cos \theta \frac{d[y^2(\theta)-y^2(\theta - \pi)]}{(-dy^2(\theta) + 1)(-dy^2(\theta - \pi) + 1)} d\theta  = 0
\end{equation}
( The last expression is obtained rewriting the integral as a sum on $\left[ -\frac{\pi}{2},\frac{\pi}{2} \right]$ plus a sum on $\left[ \frac{\pi}{2},\frac{3\pi}{2} \right]$ ; and performing a variable change ). Let us study the sign of $y^2(\theta)-y^2(\theta - \pi)$. Using (\ref{H}) we deduce that $-d\frac{y^3(\theta)}{3} + y(\theta) = H - \cos \theta$ and $-d\frac{y^3(\theta - \pi)}{3} + y(\theta - \pi) = H + \cos \theta$. From these two relations we conclude that $y(\theta - \pi) = y(\theta)$ implies $\cos(\theta) = 0$ and that $y(\theta - \pi) = - y(\theta)$ implies $H=0$. Thus if $H \neq 0$ ; $y^2(\theta)-y^2(\theta - \pi)$ does not change sign on $\left[ -\frac{\pi}{2},\frac{\pi}{2} \right]$. Moreover, on the areas a) and b1) $(-dy^2(\theta) + 1)$ does not change sign. Thus if $H \neq 0$, the argument of the last integral of (\ref{Deltax}) does not change sign and $\Delta x$ can not be zero. We thus conclude that the only solution where $x$ is a periodic function of $\theta$ is the solution corresponding to $H=0$. This solution is the one obtained from (\ref{thy},\ref{thx}) with initial conditions $y(0)=0$ and $\theta(0)=\frac{\pi}{2}$. As we have previously excluded the area $b2$ ( when $d>\frac{4}{9}$ ) we conclude that no vortex solution exists when $d>\frac{4}{9}$. We conclude that equations (\ref{thy},\ref{thx}) with $d>0$, defining the curve formed by the jet, admit one periodic solution both in $x$ and $y$ corresponding to vortices, only when $d < d_{max} = \frac{4}{9}$. This solution corresponds to $H=0$ (\ref{H}). The vortex then admits a latitudinal and a zonal axes of symmetry.

 \end{appendix}

\newpage

{\bf Figure Captions.}\\

Figure \ref{jets} : Annular jets observed in the atmosphere of Jupiter. a) Velocity field in the Great Red Spot of Jupiter (20$^0$ South), from Dowling and Ingersoll (1989). b) Velocity field in the cyclonic Barge of Jupiter (14$^0$ North) from Hatzes et al (1981) \\

Figure \ref{conint} : (a) Graphical representation of the algebraic
 equation (\ref{alg}), with the rescaled variable $\phi\equiv -\alpha/C+\psi/R^2$. The three solutions are at the intersection of the curve (left-hand side) and straight line (right-hand side). Here the integrability condition $\alpha = C_0B$ for the differential equation (\ref{jet}) \\ 

Figure \ref{figenlibre} : The free energy 
density $f(p)$ (\ref{fp}) versus the probability $p$. For $C_0 > 1$ 
and $(C_0B - \alpha_0)$ small enough $f(p)$ has two local minima and
 one local maximum, allowing to obtain two values $p_{\pm 1}$ in the 
maximization of entropy under constraints. \\

Figure \ref{figu} : The parameter $u$ versus the Lagrange parameter $C_0$, as the solution of (\ref{uu}). \\

Figure \ref{figjet} : Typical stream function profile in a jet ($u$ = 0.75 ) versus the transverse coordinate $\tau = \zeta/R$ ( Top ) and corresponding velocity profile ( Bottom ). \\

Figure \ref{lar} : Jet properties versus the segregation parameter $u$. a) Jet width, defined as the width of the region with 
velocity greater than half the maximum jet velocity. b) Maximum velocity $\left(d\phi/d\tau\right)_{max}$ and jet kinetic energy $e(u)$ ( dotted line). \\

Figure \ref{phase} : Phase diagram of the Gibbs states versus the energy $E$ and the asymmetry parameter B. The outer line is the maximum energy achievable for a
fixed B : $ E = \frac{R^2}{2}(1 - B^2)$. The frontiers line between the straight jets and the
circular jets corresponds to $A_{1} = 1/ \pi$ or $A_{-1} = 1/ \pi$. it
as been calculated using (\ref{aire}) and (\ref{eneu}) : $E =
R^2B^2(2\pi-2)/(\pi -2)^2$. The dot line represents the
  frontiers between axisymmetric vortices and the circular jets. We define it as the energy value
  for which the circular vortex area $A_{1}$ or
  $A_{-1}$ (\ref{aire}) is equal to $(2l )^2$, where $l$
  is the typical jets width ( figure \ref{lar} ).  Such a
  line depends on the numerical value of R the ratio of the Rossby
  deformation radius to the domain scale. It has been here numerically
  calculated for R = 0.03. \\

Figure \ref{intaxi} : (a) Graphical representation of the algebraic
 equation (\ref{alg}), with the rescaled
variable $\phi\equiv -\alpha/C+\psi/R^2$, like in figure \ref{conint}, but in the case
 of a Gibbs state with an
 axisymmetric vortex ($\Delta C > 0$). Then the rhs hatched area is greater than the
 lhs one. \newline (b) The corresponding potential $U(\phi)$, given by (\ref{Uaxi}), is asymmetric
 to compensate for the friction term in equation (\ref{potaxi}). \\

Figure \ref{figaxi} : Various axisymmetric stream-function profiles for decreasing $\Delta C$ ( $\Delta C = [0.9 0.6 0.3 0.1 0.05 0.03 0.01]$ and $B=0.75$. \\

Figure \ref{ellipse} : a ) Typical sub-domain shape with a topography $h(y) = ay^2$. The parameter $d$ has been chosen such that the ratio of the length on the width be 2 ; as on Jupiter's GRS. b ) Typical sub-domain shape with a topography $h(y) = ay^2$ when the parameter $d$ id very close to its maximum value $d = \frac{4}{9}$. The shape is then very elongated, with latitudinal boundaries quasi parallel, as for instance the Jovian cyclonic vortices ( 'Barges' ) described by ( Hatzes et al 1981 ). \\

Figure \ref{dimellipse} : a ) Sub-domain non dimensional length and width versus the parameter $d$ ( topography $h(y) = ay^2$ ). b ) Sub-domain aspect ratio versus the parameter $d$. \\

Figure \ref{figtopdow} : QG topography ( units $s^{-1}$ ) versus latitude computed from data of Dowling and Ingersoll (1989) : a) under the GRS ; b) under the Oval BC. \\

Figure \ref{vitMit} : Velocity profile within the GRS from ( Mitchell et al 1981 ). They have observed that the velocity is nearly tangential to ellipses. Using a grid of concentric ellipses of constant eccentricity, the velocities have been plotted with respect to the semi-major axis $\cal{A}$ of the ellipse on which the measurement point lies. Results have been fitted by a quartic in $\cal{A}$. $l^{\star}$ is the jet width, defined as the width on which the jet velocity is greater than half the maximum velocity. \\ 

Figure \ref{figaa1-R} : A) Coefficient $a$ for a quadratic topography $h(y) = ay^2$ versus $R^{\star}$ computed from our QG model (\ref{astar}). The three cross show the coefficient $a$ computed from QG topography deduced from Dowling SW observed results. 	B) Difference of PV levels $a_1-a_{-1}$ versus $R^{\star}$ computed from our QG model (\ref{temps}). The dot line represents the planetary vorticity $f_0$ at the latitude of the center of the GRS. This show that one of the PV levels may be interpreted as vorticity generated by convection-plumes from the sublayer. \\

Figure \ref{figuR} : The segregation parameter $u$ versus the Rossby deformation Radius $R^*$ for the GRS. $u$ has been computed using the actual jet maximum velocity and width ( see section (\ref{secGRS}) ). \\

Figure \ref{figdR} : The non dimensional  parameter giving the shape of the curve : $d$ ( see (\ref{thy}) ) with respect to $R^{\star}$ in our model of the GRS. The dot line represents the critical value $d = \frac{4}{9}$ below which a vortex solution exists. The ratio of the length to the width of the GRS is approximately $2$. From figure \ref{dimellipse} we conclude that this correspond to $d$ very close to the critical value $\frac{4}{9}$. From this figure, our model predicts that the Rossby deformation radius is $R^* = 1800$ km ( see section (\ref{secGRS}) for comments ). \\

Figure \ref{figly} : Definition of $l(y)$. \\

Figure \ref{Hphase} : Phase portraits of the Hamiltonian H (\ref{H}) for $y_0=0$, governing the jet shape via differential equations (\ref{thy}) ( two periods in $\theta$ ). For vortices, we are looking for periodic solutions in $y$. Thus only trajectories of areas a) and b) are under interest. Conversely trajectories of area c) could correspond to oscillating jets. The parameters $d$ governs a transition between two type of phase portraits. A) For $d<\frac{4}{9}$ ( here $d=0.075$ ), trajectories of area a) can define $y$ as a function of $\theta$ corresponding to convex vortices. B )  For $d>\frac{4}{9}$ ( here $d=0.075$ ), for trajectories of area a), the curve $y(\theta)$ admits double points. Thus they can not define vortex boundaries. \\

Figure 19 :  Resolution of the equations (\ref{frdetcha} and \ref{deltaphifr}). The long-doted and the doted lines represent the two cases discussed in appendix (\ref{appC}). \\

\newpage

\renewcommand{\floatpagefraction}{.9}
\renewcommand{\textfraction}{.1}

\begin{figure}
\caption{Annular jets observed in the atmosphere of Jupiter. a) Velocity field in the Great Red Spot of Jupiter (20$^0$ South), from Dowling and Ingersoll (1989). b) Velocity field in the cyclonic Barge of Jupiter (14$^0$ North) from Hatzes et al (1981)}
\label{jets}
\end{figure}

\newpage

\begin{figure}
\caption{(a) Graphical representation of the algebraic
 equation (\ref{alg}), with the rescaled variable $\phi\equiv -\alpha/C+\psi/R^2$. The three solutions are at the intersection of the curve (left-hand side) and straight line (right-hand side). Here the integrability condition $\alpha = C_0B$ for the differential equation (\ref{jet}) is verified, so the  two hatched
 areas are equal. \newline
b) The corresponding potential $U(\phi)$, given by (\ref{Uint}), integral from 0 to
$\phi$ of the difference between the two curves (hatched area in (a)).}
\label{conint}
  \begin{center}
    \includegraphics[width=12cm]{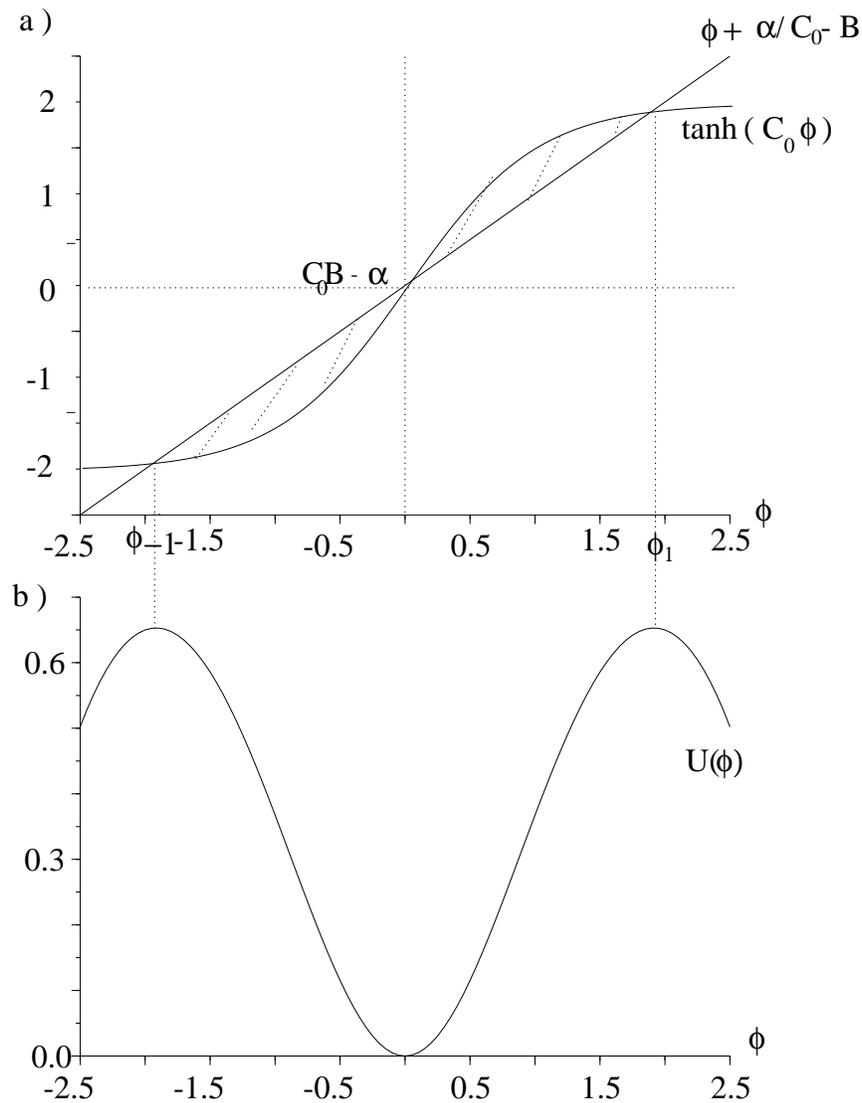}
  \end{center}
\end{figure}  

 \begin{figure}
\caption{The free energy 
density $f(p)$ (\ref{fp}) versus the probability $p$. For $C_0 > 1$ 
and $(C_0B - \alpha_0)$ small enough $f(p)$ has two local minima and
 one local maximum, allowing to obtain two values $p_{\pm 1}$ in the 
maximization of entropy under constraints.}
\label{figenlibre}
  \begin{center}
    \includegraphics[width=12cm]{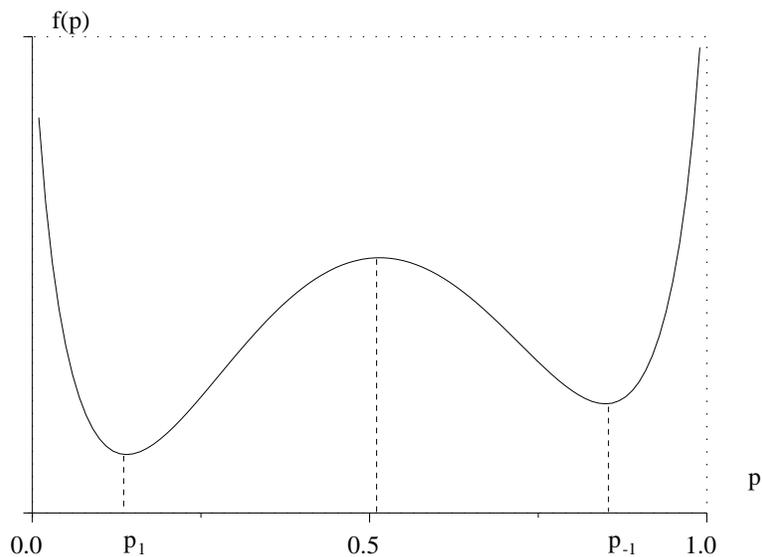}
  \end{center}\end{figure}

  \begin{figure}
  \caption{The parameter $u$ versus the Lagrange parameter $C_0$, as the solution of (\ref{uu}).}
\label{figu}
  \begin{center}
    \includegraphics[width=12cm]{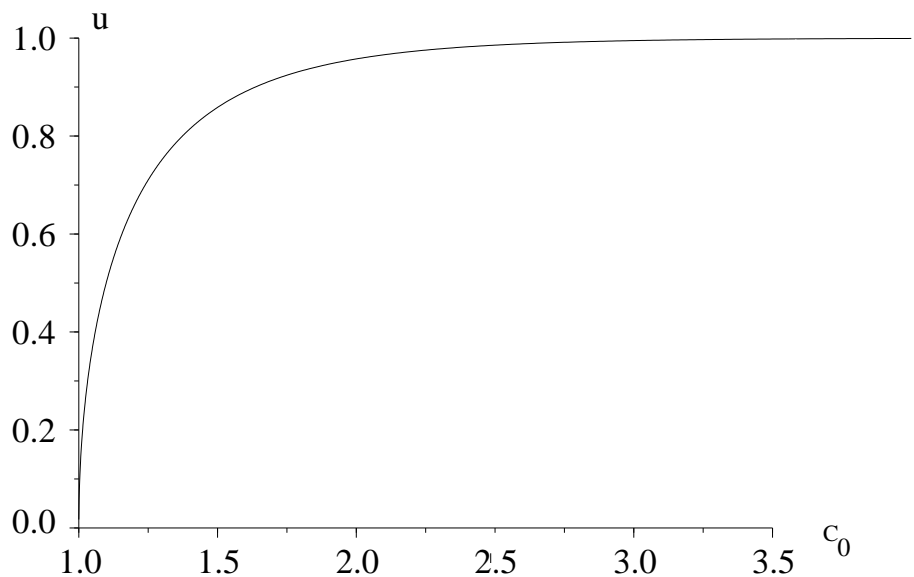}
  \end{center}
\end{figure}  
 
\begin{figure}
 
  \caption{Typical stream function profile in a jet ($u$ = 0.75 ) versus the transverse coordinate $\tau = \zeta/R$ ( Top ) and corresponding velocity profile ( Bottom )}
 \label{figjet}
  \begin{center}
    \includegraphics[width=12cm]{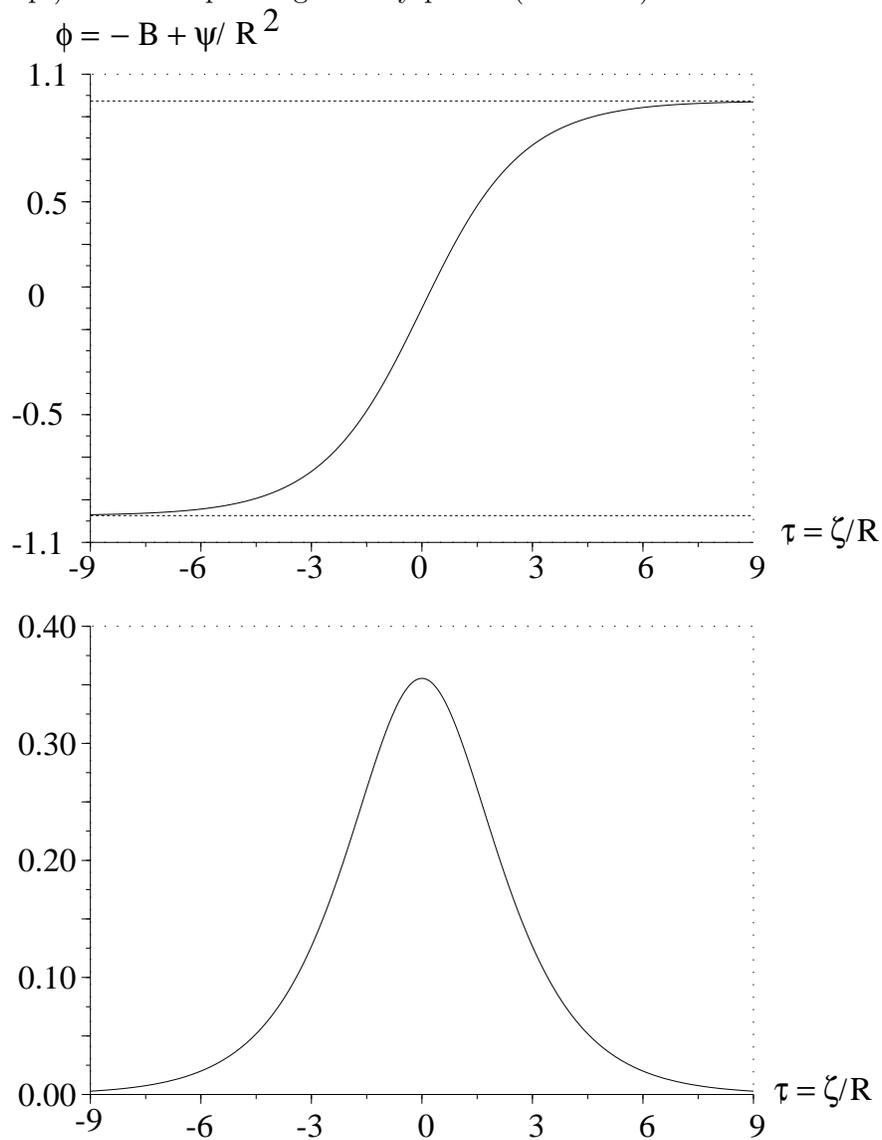}
  \end{center}
\end{figure}  
 
\begin{figure}
  \caption{Jet properties versus the segregation parameter $u$. a) Jet width, defined as the width of the region with 
velocity greater than half the maximum jet velocity. b) Maximum velocity $\left(d\phi /d\tau\right)_{max}$ and jet kinetic energy $e(u)$ ( dotted line). }
 \label{lar}
  \begin{center}
    \includegraphics[width=12cm]{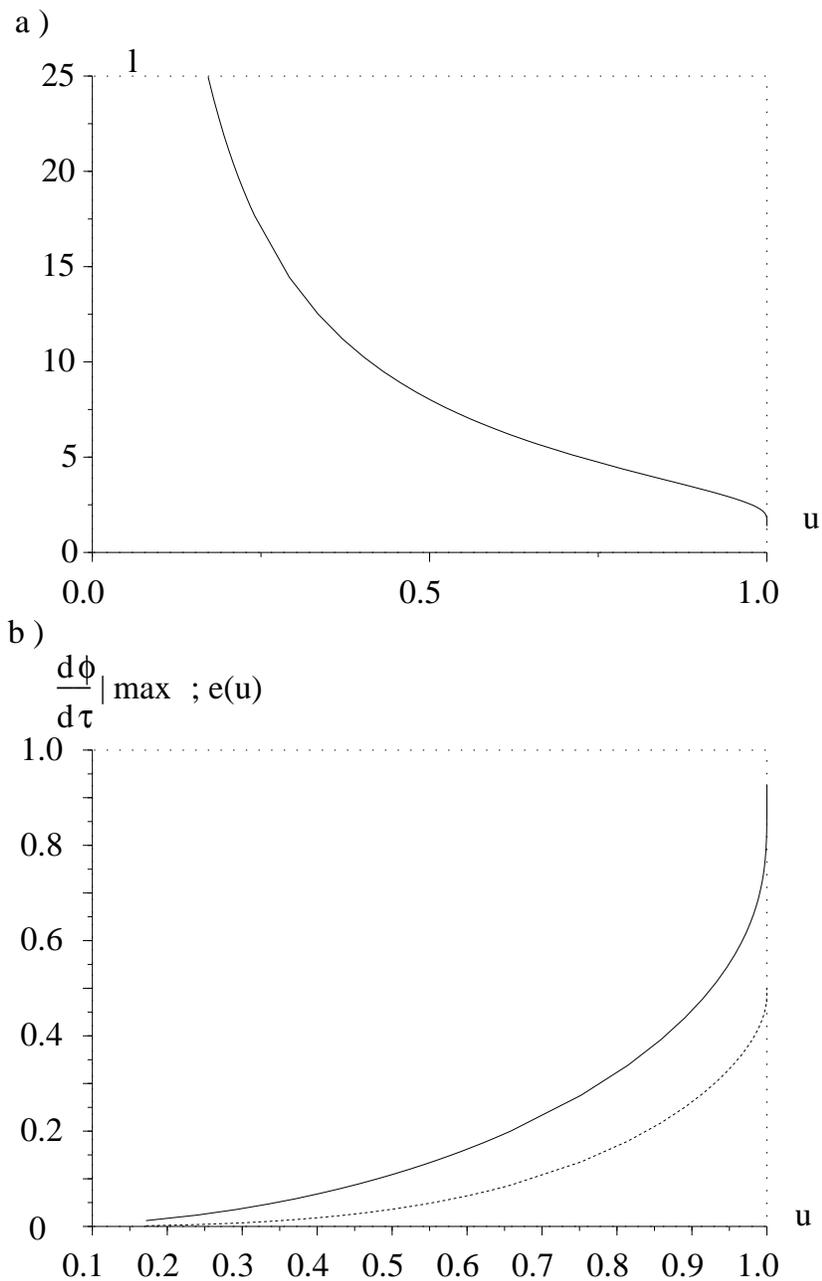}
  \end{center}
\end{figure}  
 
\begin{figure}
  \caption{Phase diagram of the Gibbs states versus the energy $E$ and the asymmetry parameter B. The outer line is the maximum energy achievable for a
fixed B : $ E = \frac{R^2}{2}(1 - B^2)$. The frontiers line between the straight jets and the
circular jets corresponds to $A_{1} = 1/ \pi$ or $A_{-1} = 1/ \pi$. it
as been calculated using (\ref{aire}) and (\ref{eneu}) :  $E =
R^2B^2(2\pi-2)/(\pi -2)^2$. The dot line represents the
  frontiers between axisymmetric vortices and the circular jets. We define it as the energy value
  for which the circular vortex area $A_{1}$ or
  $A_{-1}$ (\ref{aire}) is equal to $(2l )^2$, where $l$
  is the typical jets width ( figure \ref{lar} ).  Such a
  line depends on the numerical value of R the ratio of the Rossby
  deformation radius to the domain scale. It has been here numerically
  calculated for R = 0.03.  }
 \label{phase}
  \begin{center}
    \includegraphics[width=12cm]{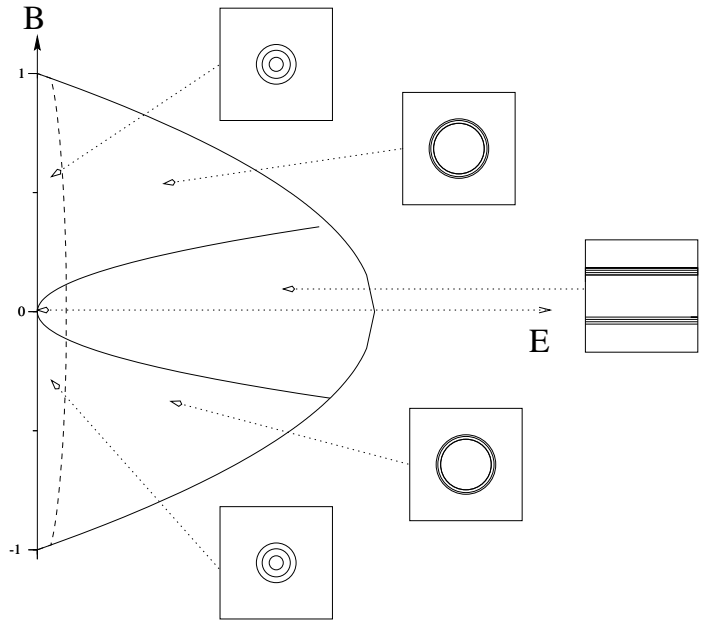}
  \end{center}
\end{figure}  

\begin{figure}
  \caption{(a) Graphical representation of the algebraic
 equation (\ref{alg}), with the rescaled
variable $\phi\equiv -\alpha/C+\psi/R^2$, like in figure \ref{conint}, but in the case
 of a Gibbs state with an
 axisymmetric vortex ($\Delta C > 0$). Then the rhs hatched area is greater than the
 lhs one. \newline (b) The corresponding potential $U(\phi)$, given by (\ref{Uaxi}), is asymmetric
 to compensate for the friction term in equation (\ref{potaxi}).}
 \label{intaxi}
  \begin{center}
    \includegraphics[width=12cm]{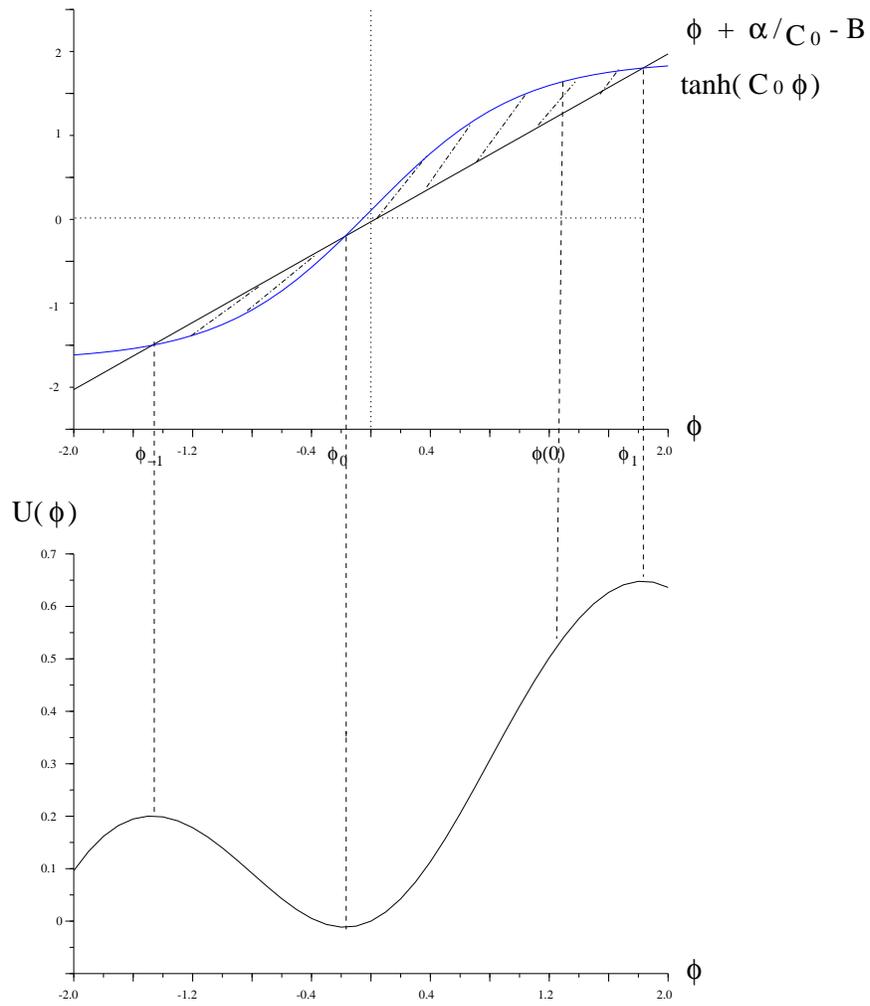}
  \end{center}
\end{figure}  
 
\begin{figure}
\caption{Various axisymmetric stream-function profiles for decreasing $\Delta C$ ( $\Delta C = [0.9 0.6 0.3 0.1 0.05 0.03 0.01]$ and $B=0.75$. }
\label{figaxi}
\begin{center}
\includegraphics[width=12cm]{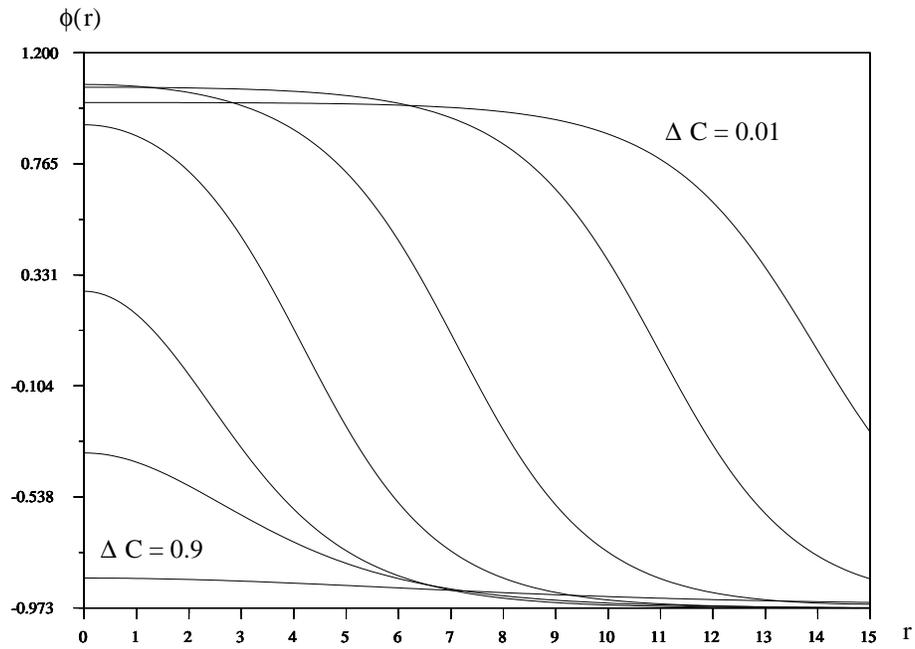}
\end{center}
\end{figure}  
 

\begin{figure} 
\caption{a ) Typical sub-domain shape with a topography $h(y) = ay^2$. The parameter $d$ has been chosen such that the ratio of the length on the width be 2 ; as on Jupiter's GRS. b ) Typical sub-domain shape with a topography $h(y) = ay^2$ when the parameter $d$ id very close to its maximum value $d = \frac{4}{9}$. The shape is then very elongated, with latitudinal boundaries quasi parallel, as for instance the Jovian cyclonic vortices ( 'Barges' ) described by ( Hatzes et al 1981 ).} 
\label{ellipse}  
\begin{center}    
\includegraphics[width=12cm]{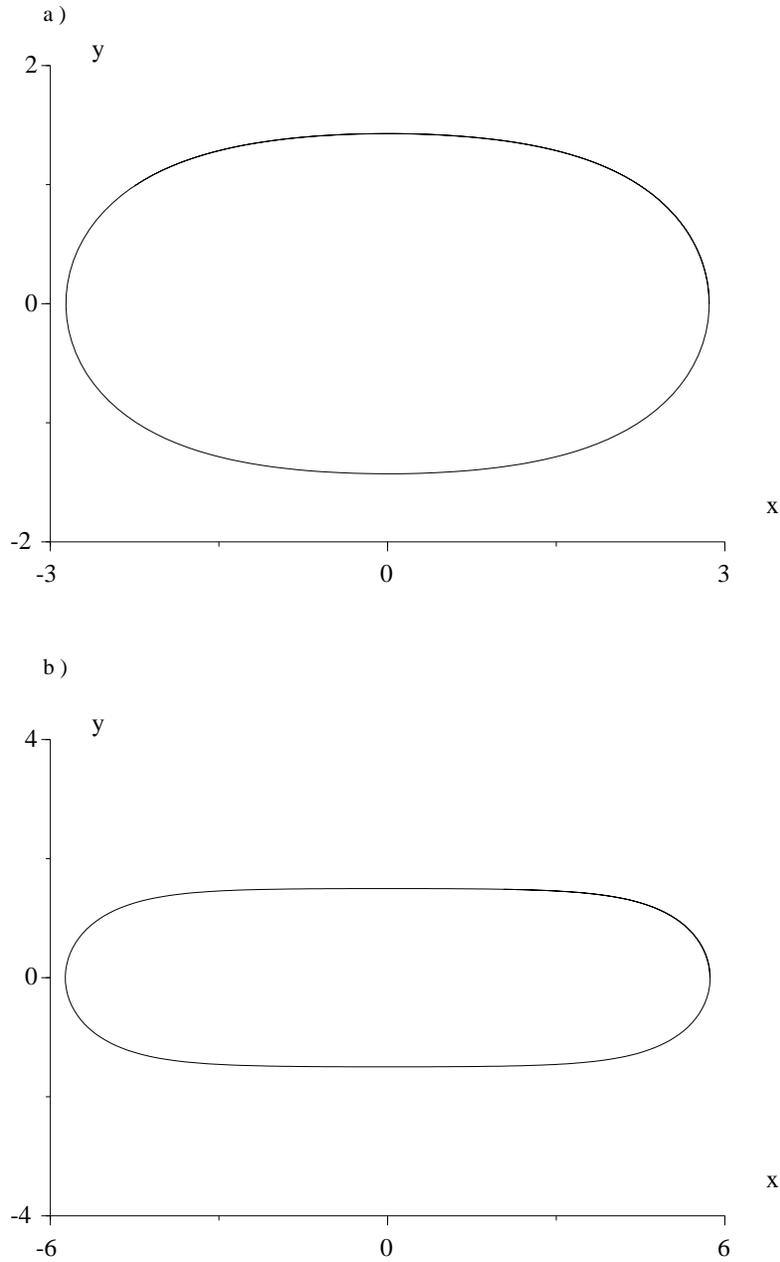}  
\end{center}
\end{figure} 

\begin{figure}  
\caption{a ) Sub-domain non dimensional length and width versus the parameter $d$ ( topography $h(y) = ay^2$ ). b ) Sub-domain aspect ratio versus the parameter $d$. }  
\label{dimellipse}
\begin{center}
\includegraphics[width=12cm]{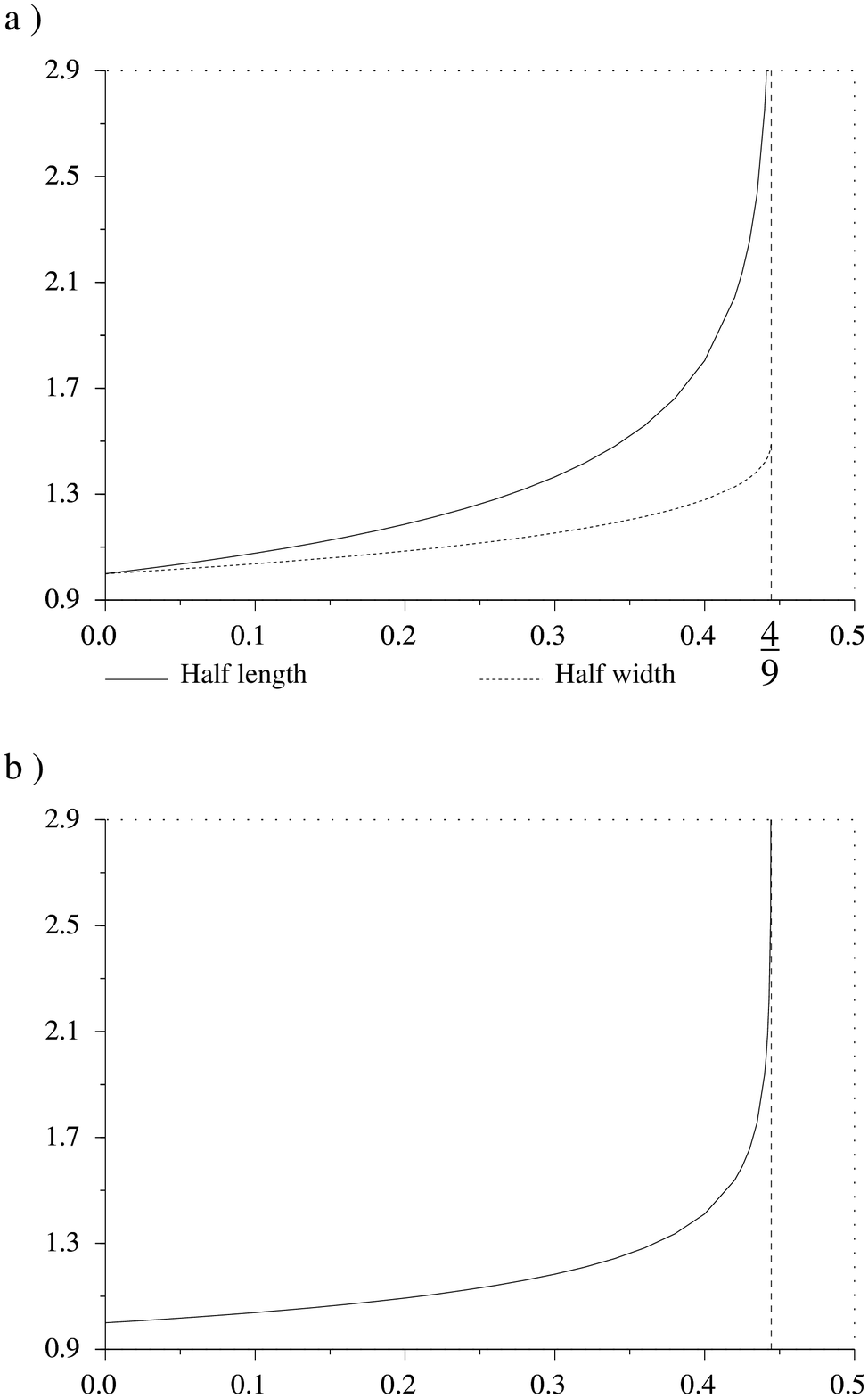}
\end{center}
\end{figure}

\begin{figure}  
\caption{QG topography ( units $s^{-1}$ ) versus latitude computed from data of Dowling and Ingersoll (1989) : a) under the GRS ; b) under the Oval BC.} 
\label{figtopdow} 
\begin{center}  
\includegraphics[width=12cm]{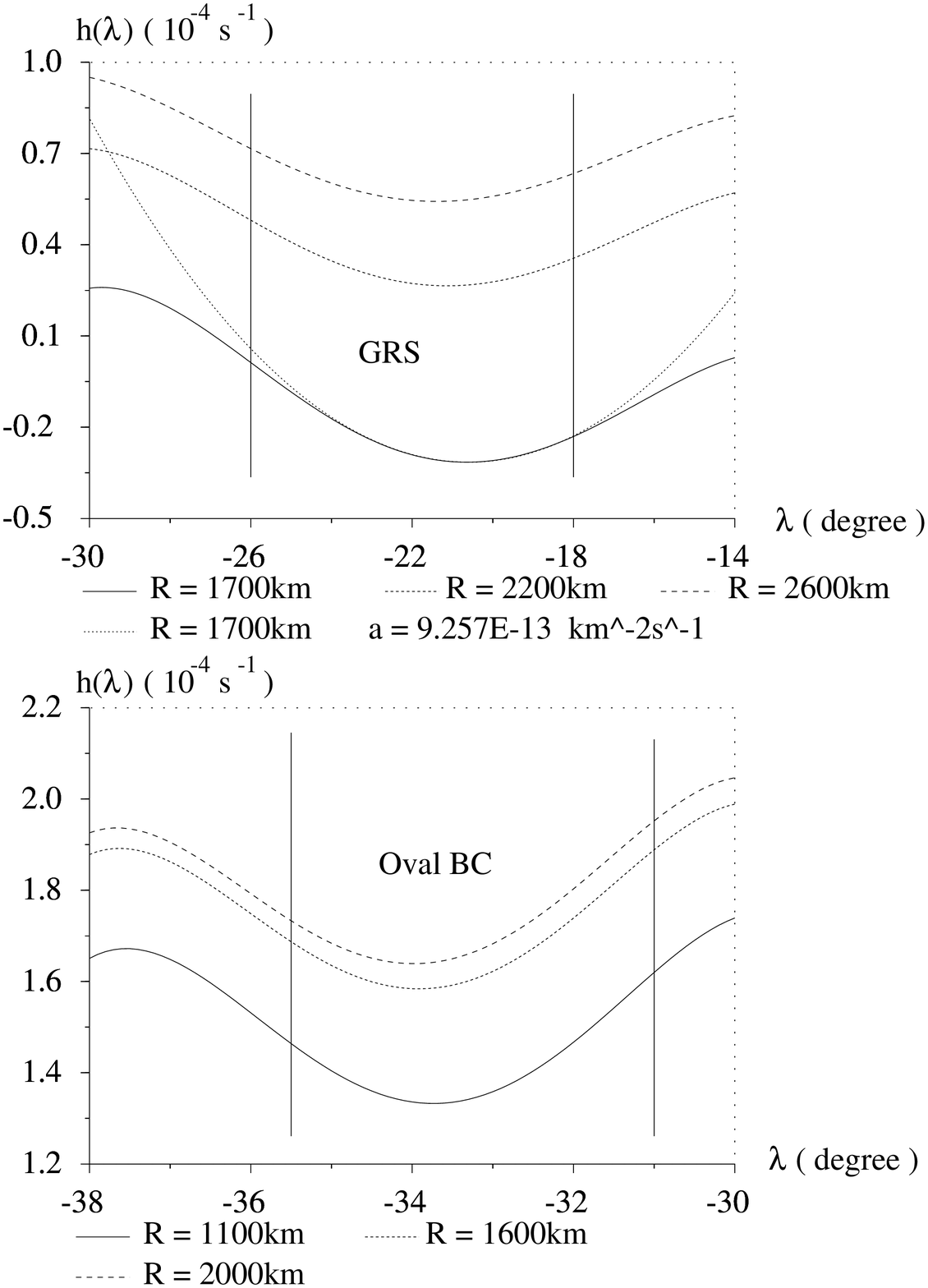} 
\end{center}
\end{figure}

\begin{figure}  
\caption{Velocity profile within the GRS from ( Mitchell et al 1981 ). They have observed that the velocity is nearly tangential to ellipses. Using a grid of concentric ellipses of constant eccentricity, the velocities have been plotted with respect to the semi-major axis $\cal{A}$ of the ellipse on which the measurement point lies. Results have been fitted by a quartic in $\cal{A}$. $l^{\star}$ is the jet width, defined as the width on which the jet velocity is greater than half the maximum velocity.} 
\label{vitMit}  
\begin{center}    
\includegraphics[width=12cm]{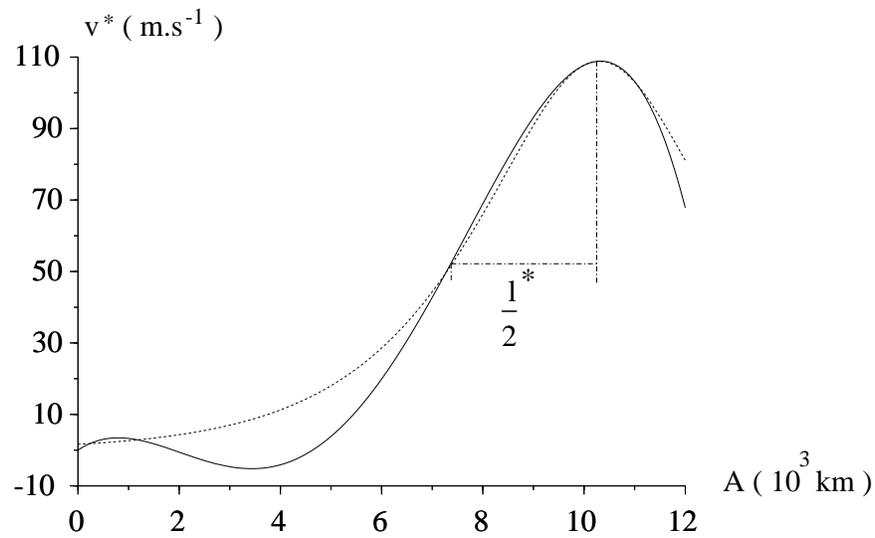}
\end{center}
\end{figure}  

\begin{figure}  
\caption{ A) Coefficient $a$ for a quadratic topography $h(y) = ay^2$ versus $R^{\star}$ computed from our QG model (\ref{astar}). The three cross show the coefficient $a$ computed from QG topography deduced from Dowling SW observed results. 	B) Difference of PV levels $a_1-a_{-1}$ versus $R^{\star}$ computed from our QG model (\ref{temps}). The dot line represents the planetary vorticity $f_0$ at the latitude of the center of the GRS. This show that one of the PV levels may be interpreted as vorticity generated by convection-plumes from the sublayer.}  
\label{figaa1-R}  
\begin{center}    
\includegraphics[width=12cm]{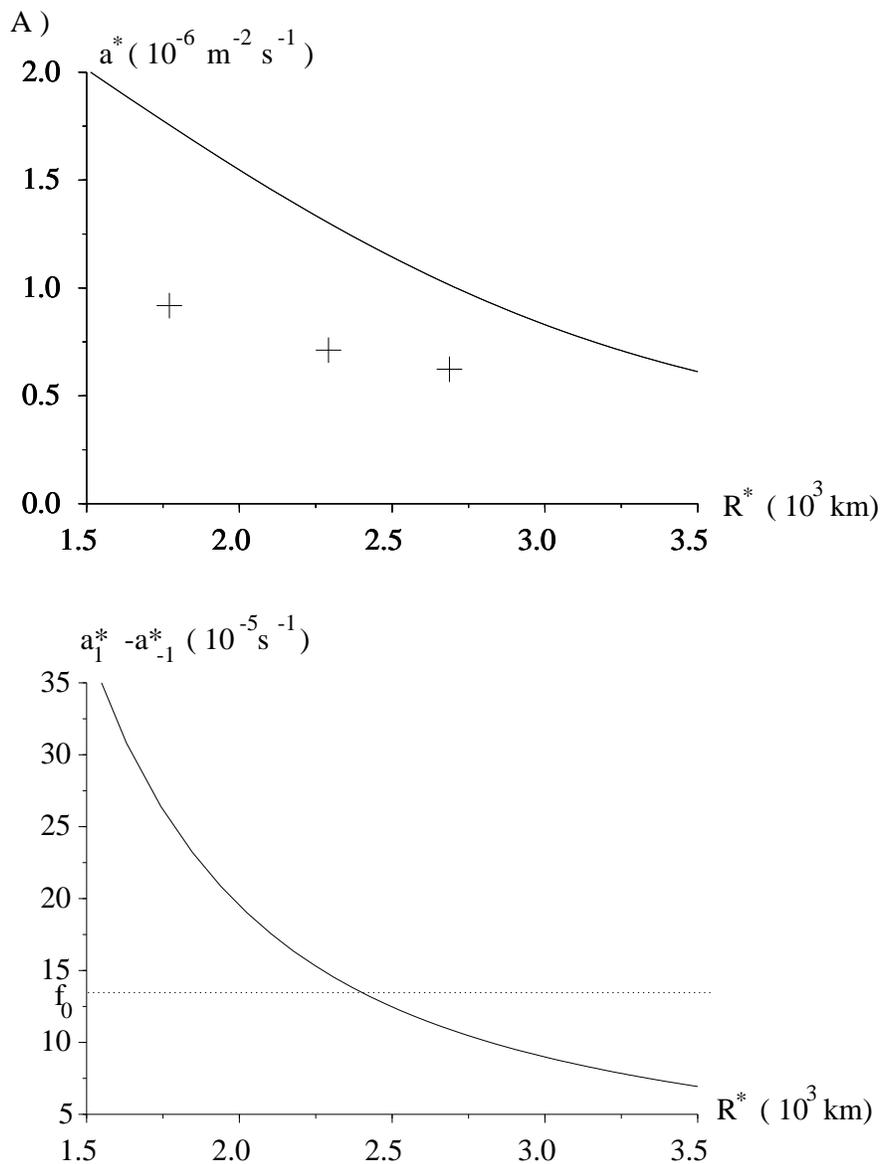}
  \end{center}
\end{figure} 

\begin{figure}  
\caption{The segregation parameter $u$ versus the Rossby deformation Radius $R^*$ for the GRS. $u$ has been computed using the actual jet maximum velocity and width ( see section (\ref{secGRS}) ).}  
\label{figuR} 
\begin{center}  
\includegraphics[width=12cm]{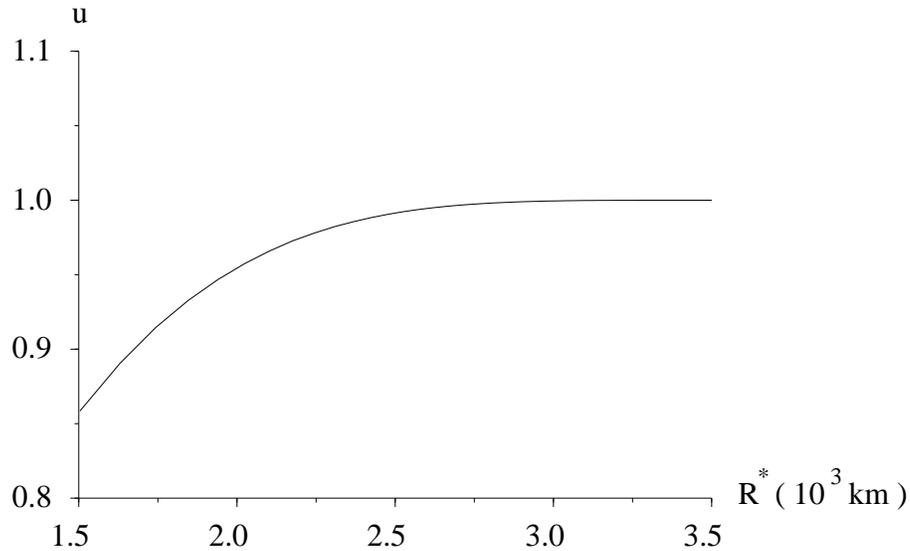}
\end{center}
\end{figure}

 \begin{figure}  
\caption{The non dimensional  parameter giving the shape of the curve : $d$ ( see (\ref{thy}) ) with respect to $R^{\star}$ in our model of the GRS. The dot line represents the critical value $d = \frac{4}{9}$ below which a vortex solution exists. The ratio of the length to the width of the GRS is approximately $2$. From figure \ref{dimellipse} we conclude that this correspond to $d$ very close to the critical value $\frac{4}{9}$. From this figure, our model predicts that the Rossby deformation radius is $R^* = 1800$ km ( see section (\ref{secGRS}) for comments ).}  
\label{figdR}  
\begin{center}    
\includegraphics[width=12cm]{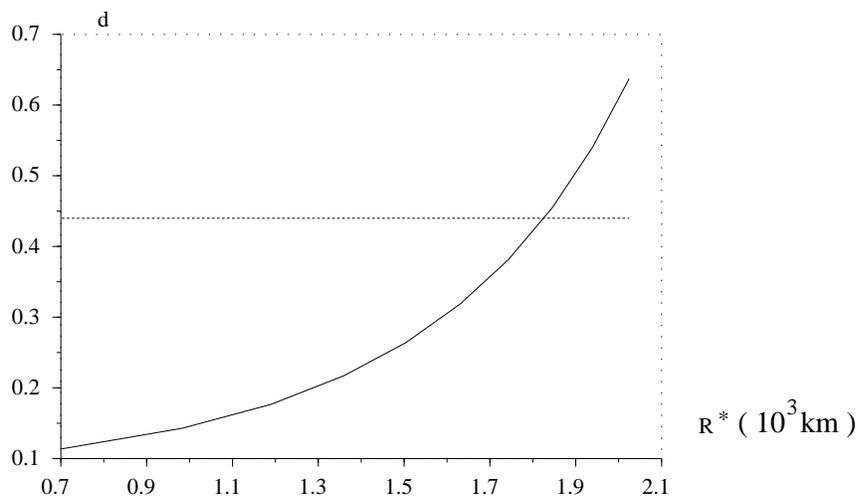}
\end{center}
\end{figure}  

\begin{figure}  
\caption{Definition of $l(y)$.}  
\label{figly}  
\begin{center}    
\includegraphics[width=12cm]{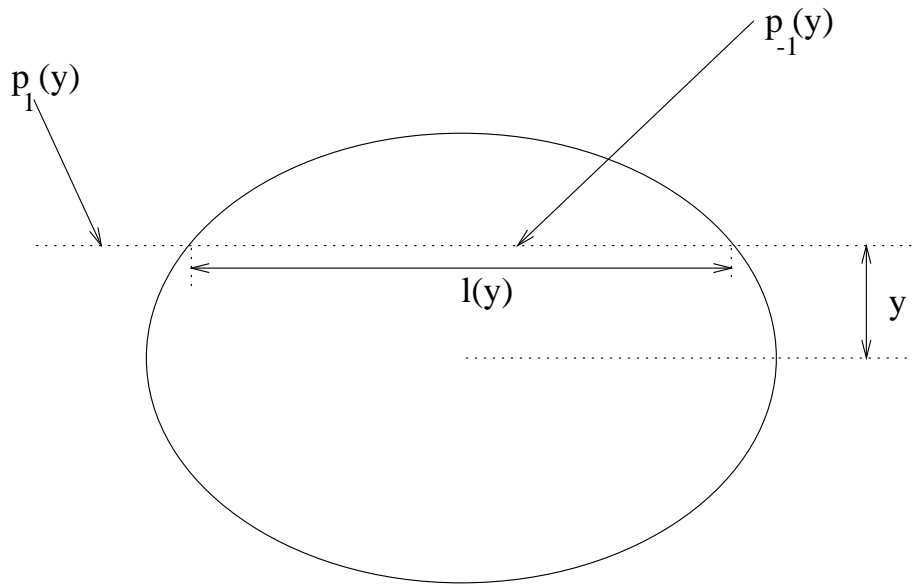}
\end{center}
\end{figure} 

\begin{figure}
\caption{Phase portraits of the Hamiltonian H (\ref{H}) for $y_0=0$, governing the jet shape via differential equations (\ref{thy}) ( two periods in $\theta$ ). For vortices, we are looking for periodic solutions in $y$. Thus only trajectories of areas a) and b) are under interest. Conversely trajectories of area c) could correspond to oscillating jets. The parameters $d$ governs a transition between two type of phase portraits. A) For $d<\frac{4}{9}$ ( here $d=0.075$ ), trajectories of area a) can define $y$ as a function of $\theta$ corresponding to convex vortices. B )  For $d>\frac{4}{9}$ ( here $d=0.075$ ), for trajectories of area a), the curve $y(\theta)$ admits double points. Thus they can not define vortex boundaries.} 
\label{Hphase}
\begin{center}    
\includegraphics[width=12cm]{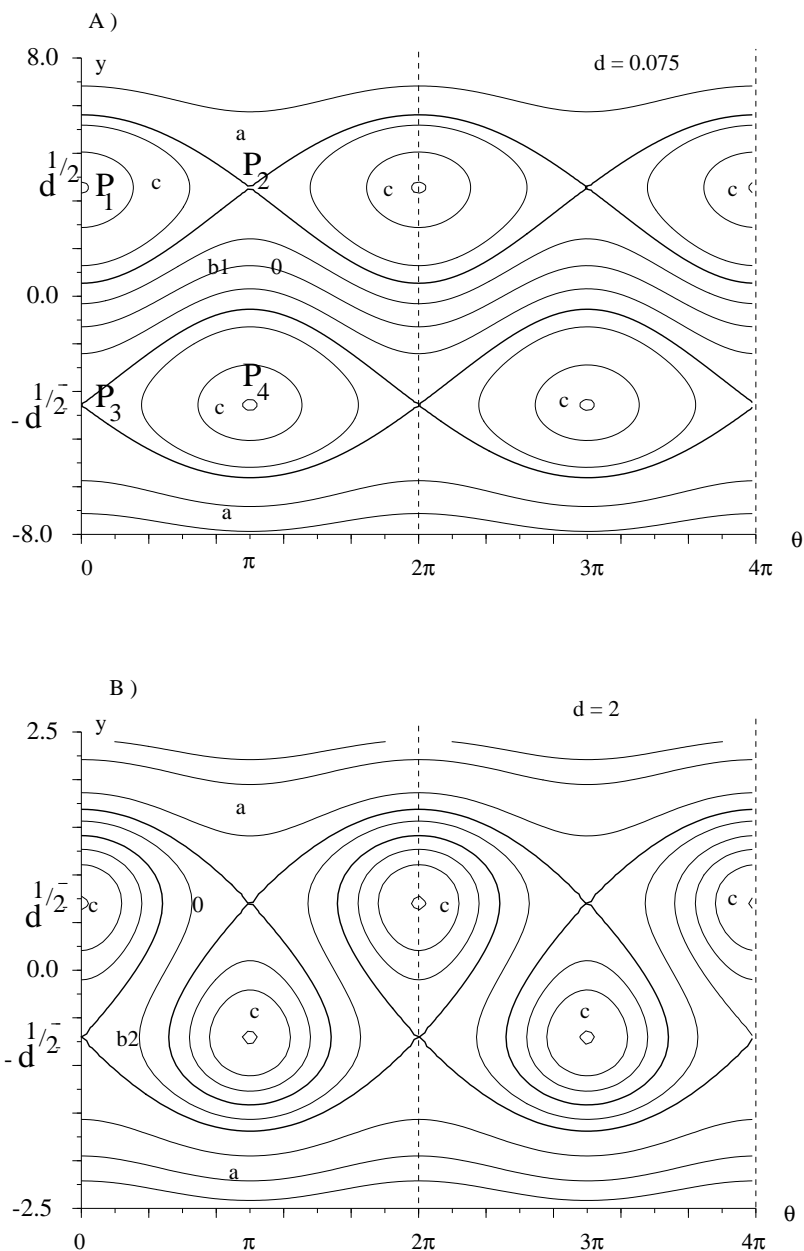}
\end{center}
\end{figure} 

\newpage
. 
\newpage  

\begin{figure}
\caption{Resolution of the equations (\ref{frdetcha} and \ref{deltaphifr}). The long-doted and the doted lines represent the two cases discussed in appendix (\ref{appC}). }
\label{potphicha}
  \begin{center}
    \includegraphics[width=12cm]{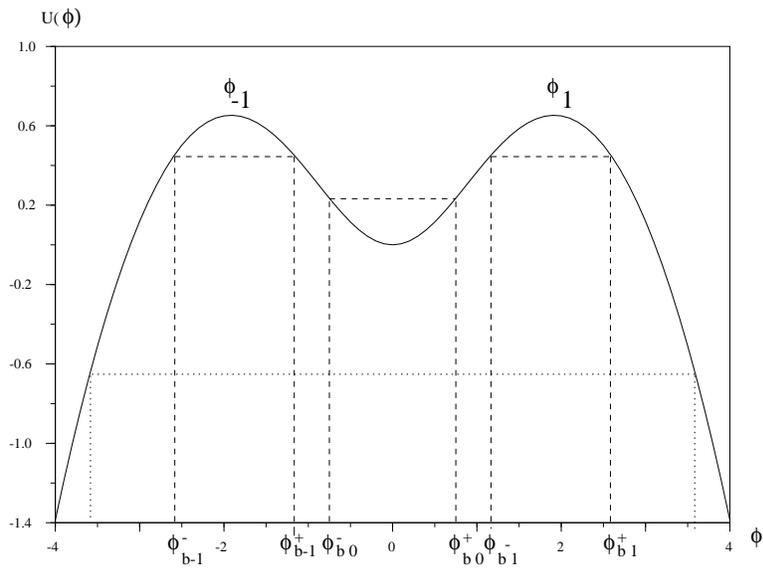}
  \end{center}
\end{figure}

\end{document}